\documentclass[12pt]{iopart}
\usepackage{iopams}
\usepackage{graphicx}

\newcommand{\CD}{{\cal D}}
\newcommand{\CR}{{\cal R}}
\newcommand{\average}[1]{\left\langle #1 \right\rangle_\CD}

\newcommand{\gaverage}[1]{\left\langle #1 \right\rangle_{\Sigma}}
\newcommand{\initial}[1]{{#1_{\rm \bf i}}}
\newcommand{\now}[1]{{#1_{\rm \bf 0}}}
\newcommand{\inI}{{I}}
\newcommand{\inII}{{II}}

\begin{document}
\title[Backreaction problem and morphon field]{Correspondence between 
kinematical backreaction and scalar field cosmologies --- the `morphon field'}

\author{Thomas Buchert}
\address{Fakult\"at f\"ur Physik, Universit\"at Bielefeld, Postfach 100131,
33501 Bielefeld, and}
\address{Arnold Sommerfeld Center for Theoretical Physics ASC,
Ludwig--Maximilians--Universit\"{a}t, Theresienstra{\ss}e 37,
80333 M\"{u}nchen, Germany \\Email: buchert@theorie.physik.uni-muenchen.de}

\author{Julien Larena and Jean--Michel Alimi}
\address{Laboratoire de l'Univers et ses Th\'eories LUTH, CNRS UMR 8102,\\
Observatoire de Paris--Meudon and Universit\'e Paris 7, 
92195 Meudon, France\\
Emails: Julien.Larena@obspm.fr and Jean-Michel.Alimi@obspm.fr}

\begin{abstract}
Spatially averaged inhomogeneous cosmologies in classical general relativity can  
be written in the form of effective Friedmann equations with sources that 
include backreaction terms. In this paper we propose to describe these backreaction 
terms with the help of a homogeneous scalar field evolving in a potential; we call it 
the `morphon field'. This new field links classical inhomogeneous cosmologies to 
scalar field cosmologies, allowing to reinterpret, e.g., quintessence scenarios by 
routing the physical origin of the scalar field source to inhomogeneities in the Universe. 
We investigate a one--parameter family of scaling solutions to the 
backreaction problem. Subcases of these solutions (all without an assumed cosmological 
constant) include scale--dependent models with Friedmannian kinematics
that can mimic the presence of a cosmological constant or a time--dependent 
cosmological term. We explicitly reconstruct the scalar field potential for the 
scaling solutions, and discuss those cases that provide a solution to the Dark Energy 
and coincidence problems. In this approach, Dark Energy emerges from morphon fields, 
a mechanism that can be understood through the proposed correspondence: 
the averaged cosmology is characterized by a weak decay (quintessence) or growth 
(phantom quintessence) of kinematical fluctuations, fed by `curvature energy' 
that is stored in the averaged 3--Ricci curvature. We find that the late--time 
trajectories of those models approach attractors that lie in the future of a state 
that is predicted by observational constraints. 
\end{abstract}


\pacs{04.20.-q, 04.20.-Cv, 04.40.-b, 95.30.-k, 95.36.+x, 98.80.-Es, 98.80.-Jk}

\section{Introduction}

The fact that the spatially averaged inhomogeneous Universe does not evolve as  
the standard model of cosmology, furnished by a 
homogeneous--isotropic solution of Einstein's laws of gravity, has recently become
a major topic aiming at a possible solution to the Dark Energy 
problem  \cite{wetterich}, \cite{beanetal:darkenergy}, \cite{straumann}. 
While the 
averaging problem in relativistic cosmology has a long history, initiated by George Ellis
\cite{ellis} (references may be found in \cite{ellisbuchert} and \cite{buchert:static}),
the backreaction terms, i.e. those averaged contributions that lead to deviations from a
Friedmannian cosmology, have been considered quantitatively unimportant.
Recent work that conjectures a large backreaction effect (e.g. \cite{kolbetal})
has been accompanied both by counter conjectures (e.g. \cite{wald}) and 
special solutions of the averaged Einstein equations \cite{buchert:grgdust} 
that support the claim that backreaction
could be made responsible for the Dark Energy gap (e.g. \cite{buchert:darkenergy}, 
\cite{nambu}, \cite{alnes:LTB}, 
\cite{rasanen:constraints,rasanen:model}, \cite{curvatureLTB} 
and references and discussion in \cite{ellisbuchert}).

In parallel to this discussion, other models have been extensively investigated, most notably 
quintessence models that invoke the presence of a scalar field source, this
scalar field being a standard one or a phantom field (with a negative kinetic energy)
\cite{Wetterich88,Peebles2003,Faraoni2005,Jassal2006} and references in
the reviews \cite{sahnistarobinskii} and \cite{copeland}. 
Those models imply the possibility of a scenario featuring a time--dependent cosmological term.
Particular properties of the scalar field potential are discussed on phenomenological
grounds, which can lead to an acceleration of the regional Universe and so to an 
explanation of Dark Energy. A further problem in conjunction with modeling a repulsive 
component in universe models is known as the {\it coincidence problem}, i.e. a recent domination
of this component is favoured, possibly occuring around the epoch of structure formation.
Its solution also motivates the construction of models with a time--dependent
cosmological term that needs no `fine--tuning'.

In this paper we also consider scalar field cosmologies, but we shift the perspective from the 
usual interpretation of a scalar field {\it source} in Einstein's equations to a mean field description
of averaged inhomogeneities. We see a number of advantages entailed by such a description
that motivated the present investigation.
First, a (homogeneous) scalar field as a model of spatially averaged (i.e. effectively homogeneous)
geometrical degrees of freedom that are physically present does not need to be justified as 
an additional source arising from fundamental field theories; it has a
well--defined physical status and, as such, does not suffer from a phenomenological 
parameterization, since it is constrained by Einstein's field
equations. Second, inhomogeneities encoded in backreaction terms {\it give rise to} 
the scalar field cosmology, fixing  its potential, its equation of state, etc.; they not 
only influence the evolution of the scalar field, but they determine it. 
Third, the proposed correspondence allows a realistic reinterpretation of 
quintessence models and other phenomenological approaches involving scalar fields.
These approaches  must not be considered as 
independent alternatives, they here describe the same physics that also underlies the backreaction
approach, and so can be confronted with physical constraints; in other words, if backreaction
can indeed be made responsible for the late--time acceleration of the Universe, then
the effort spent on quintessence models and other approaches \cite{copeland} 
can be fruitfully exploited in terms of the proposed rephrasing. 
Fourth, since the underlying effective equations are general and do not invoke
perturbative assumptions, the scalar field cosmology provides access to the whole
solution space, notably to the non--perturbative regime.
Finally,  subcases of exact solutions include a cosmological constant, which therefore
can or cannot be included in Einstein's equations; the cosmological constant is a 
particular solution of the averaged inhomogeneous model without necessarily being present in 
Einstein's equations.

Since the scalar field introduced in this way stems from averaged geometrical degrees of freedom,
i.e. kinematical backreaction terms encoded in the extrinsic curvature of spatial hypersurfaces,
as well as their scalar curvature, we call it the {\it morphon field}: it effectively models
the  inhomogeneities {\it shaping} the structure in the Universe and capturing the total effect 
of kinematical backreaction, i.e. the effect leading to a deviation from the kinematics of the
standard model.  Speaking in terms of geometry, we may roughly say that the morphon models the 
geometrical vacuum degrees of freedom in Einstein's theory, which are neglected when 
modeling the Universe with the help of averaged matter sources only.
We call the scalar field a {\it morphon} only if it arises from the proposed mean field 
description; we can of course still consider other scalar fields as sources of Einstein's equations.
We expect, and we shall demonstrate this for quintessence fields, 
that the {\it morphon} is capable of acting like any other 
scalar field model,  e.g., assigned to an inflaton, a curvaton, a dilaton, etc., 
depending on our capability to exploit the proposed correspondence, i.e. the 
possibility to find  solutions of the averaged Einstein equations and to reconstruct the
scalar field potential for this solution. 

We here provide a first step that should
help to appreciate the various possibilities, but this investigation does not claim to be 
exhaustive. In this line we are going to choose a minimal parameterization
of the scalar field cosmology in the following sense: first, we do not include the cosmological
constant and the constant curvature term in the `Friedmannian part' of the averaged 
equations, since both arise as subcases of special solutions for the backreaction terms; 
second, we choose the simplest foliation (constant lapse and
vanishing shift in the ADM (Arnowitt Deser Misner)
formulation of Einstein's equations) and the simplest matter
model `irrotational dust' in order to exemplify the correspondence; third, we realize the
correspondence for a particular class of scaling solutions with single--power laws. 
In this framework we show that the correspondence holds with a standard minimally
coupled scalar field that can play the role of a quintessence field. If we include the full
coordinate degrees of freedom as well as an inhomogeneous pressure source in a perfect fluid
energy momentum tensor (or even imperfect fluid sources), 
more sophisticated scalar field cosmologies would arise.
We comment on this possibility at the end of the paper.  

The proposed correspondence can also be inverted, e.g. we may start with a known
model of quintessence and try to recover the corresponding solution of the averaged
Einstein equations. We shall  not exemplify
this inversion in the present paper, we just note that free parameters in a given 
quintessence model can be determined through this correspondence.
 
We proceed as follows. Section 2 recalls the basic equations and relations needed for
the present work. Section 3 sets up the correspondence between kinematical
backreaction and scalar field models. Section 4 investigates a family of  scaling solutions
to the backreaction problem. 
Section 5 exploits the proposed correspondence. 
Here we explicitly reconstruct the potential for the scaling solutions, 
explore the solution space of inhomogeneous
cosmologies with the help of the scaling solutions, and discuss  
some particular cases, which have been advanced before in the literature, 
and which are now interpreted with the help of a morphon field. 
We also give a concrete model and confront it with observational constraints. 
Section 6 contains a summary and an outlook on possible generalizations.

\section{The backreaction context}

The averaging problem in relativistic cosmology involves a variety of approaches and problems.
Generally, research work in this field deals with averaging inhomogeneities in matter and/or
geometry. Mostly, spatial averaging is envisaged, but also
averaging on the light--cone is accessible by some averaging techniques (compare discussions
and references in \cite{ellis}, \cite{ellisbuchert}, \cite{carfora:RG,klingon,buchertcarfora:PRL}.)

In this work we focus on the comparatively simple approach of averaging the scalar parts
of Einstein's equations on a given foliation of spacetime.
The time--evolution of integral properties of the cosmological model 
on compact spatial domains can be extracted from Einstein's equation 
without any perturbative assumptions by Riemannian volume integration. 
The simplest example is the time--evolution of the volume.
Thus, we do not aim at changing the physics of the inhomogeneous
cosmological model. More ambitious and physically different in motivation are 
averaging strategies  that effectively replace the inhomogeneous 
hypersurfaces and inhomogeneous tensor fields on these hypersurfaces by another,
smoothed universe model \cite{klingon}, \cite{zala05}. These latter techniques
lead to further `intrinsic backreaction' effects by, e.g. flowing averages on a bumpy 
geometry to averages on a constant--curvature `template universe' as a fitting device.
This can be nicely put into practice using the Ricci--Hamilton flow that renormalizes
the averaged variables and leads to a `dressing' of cosmological parameters
by those additional `backreaction' terms
\cite{klingon},\cite{buchertcarfora:PRL}.
 
With kinematical averaging we  aim at an effective description of the kinematics of
the inhomogeneous Universe, and we still encounter a number of scalar contributions 
that add up to the 
averaged matter sources in an inhomogeneous model (`kinematical backreaction terms')
as a result of the fact that spatial averaging and time--evolution are non--commuting operations
(not necessarily to be attributed to the nonlinearity of Einstein's equations).
Employing a perfect fluid source in the energy momentum tensor, those `backreaction 
terms' comprise
a contribution from averaged expansion and shear fluctuations (i.e., terms encoding extrinsic
curvature of the hypersurfaces), 
the averaged 3--Ricci curvature (i.e,  a term that encodes intrinsic curvature of
the hypersurfaces), an averaged pressure gradient (or an averaged acceleration divergence), and 
frame fluctuation terms, i.e. coordinate effects (like the averaged variance of the lapse 
function in the ADM  setting). 
In this framework and for vanishing shift, but arbitrary lapse and arbitrary 3--metric, 
the general equations were given  in \cite{buchert:grgfluid}.
However, we further restrict the analysis to a universe model filled with an irrotational fluid of 
dust matter \cite{buchert:grgdust} in order to provide the most transparent framework for
the purpose of setting up the correspondence with a scalar field cosmology. 

\subsection{Averaged ADM equations for constant lapse and vanishing shift}

We employ a foliation of  spacetime into flow--orthogonal hypersurfaces with the 3--metric 
$g_{ij}$ in the line--element $ds^2 = -dt^2 + g_{ij}\,dX^i dX^j$. 
We define the following averager, restricting 
attention to scalar functions $\Psi (t,X^i )$:
\begin{equation}
\label{eq:average-GR}
\average{\Psi (t, X^i)}:=
\frac{1}{V_\CD}\int_\CD \Psi (t, X^i)\;J d^3 X  \;\;;\;\;V_\CD = \int_\CD J d^3 X \;\;,
\end{equation}
with  $J:=\sqrt{\det(g_{ij})}$; $g_{ij}$ is  an arbitrary metric  of the
spatial  hypersurfaces, and  $X^i$ are  coordinates that  are constant
along  flow lines, which are here spacetime geodesics. 
Defining a {\it volume scale factor} by the volume $V_\CD$ of a simply--connected domain 
$\CD$ in a t--hypersurface, normalized by the  volume of the initial domain
$\initial\CD$,
\begin{equation}
\label{scalefactor}
a_\CD := \left(\frac{V_\CD}{V_{\initial\CD}}\right)^{1/3}\;\;,
\end{equation}
the following exact equations can be derived \cite{buchert:grgdust}
(an overdot denotes partial time--derivative). First, by averaging {\it
Raychaudhuri's equation}
we obtain: 
\begin{equation}
\label{averageraychaudhuri}
3\frac{{\ddot a}_\CD}{a_\CD} + 4\pi G \langle\varrho\rangle_\CD - \Lambda
= {\cal Q}_\CD\;\;\;;\;\;
 \langle\varrho\rangle_\CD = \frac{M_\CD}{V_{\initial\CD}a_\CD^3}\;\;.
\end{equation}
The first integral of the above equation is directly given by averaging 
the {\it Hamiltonian constraint}:
\begin{equation}
\label{averagehamilton}
\left( \frac{{\dot a}_\CD}{a_\CD}\right)^2 - \frac{8\pi G}{3}\langle\varrho\rangle_\CD
- \frac{\Lambda}{3}=- \frac{\average{\CR} +{\cal Q}_\CD}{6} \;\;,
\end{equation}
where   the  total restmass   $M_\CD$,   the  averaged   spatial  3--Ricci   scalar
$\average{\CR}$   and   the  {\it kinematical backreaction  term}  ${\cal Q}_\CD$  
are domain--depen\-dent and, except the mass, time--depen\-dent functions.
The backreaction source term is given by
\begin{equation}
\label{eq:Q-GR} 
{\cal Q}_\CD : = 2 \average{\inII} - \frac{2}{3}\average{\inI}^2 =
\frac{2}{3}\average{\left(\theta - \average{\theta}\right)^2 } - 
2\average{\sigma^2}\;\; ;
\end{equation}
here,  $\inI = \Theta^i_{\;i}$  and $\inII = \frac{1}{2}[\,(\Theta^i_{\;i})^2 - 
\Theta^i_{\;j}\Theta^j_{\;i}\,]$  
denote  the  principal scalar invariants  of the  expansion
tensor, defined  as minus the extrinsic
curvature  tensor $K_{ij}:=-\Theta_{ij}$. In the second equality above it was split 
into kinematical invariants through the decomposition  
$\Theta_{ij} = \frac{1}{3}g_{ij}\theta + \sigma_{ij}$, with the rate of expansion 
$\theta =\Theta^i_{\;i}$, the shear tensor $\sigma_{ij}$, and the rate of shear
$\sigma^2 :=1/2 \sigma_{ij}\sigma^{ij}$;
note that vorticity is absent in the present foliation; 
we adopt the summation convention.

The time--derivative of the averaged Hamiltonian constraint (\ref{averagehamilton})
agrees with the averaged Raychaudhuri equation (\ref{averageraychaudhuri}) by virtue
of the
following {\it integrability  condition}:
\begin{equation}
\label{integrability}
\partial_t {\cal Q}_\CD + 6 H_{\cal D} {\cal Q}_\CD +  
\partial_t \average{\CR}+ 2 H_{\cal D} \average{\CR} = 0 \;\;,
\end{equation}
where we have introduced a {\it volume Hubble functional} $H_{\cal D}:= 
{\dot a}_\CD / a_\CD$.
The above equations can formally be recast into standard
Friedmann equations for an effective perfect fluid energy momentum tensor
with new effective sources \cite{buchert:grgfluid}\footnote{Note that in this
representation
of the effective equations ${p}_{\rm eff}$  denotes
an `effective pressure'; there is no pressure due to a matter source here.}:
\begin{equation}
\label{equationofstate}
\fl
\varrho^{\CD}_{\rm eff} = \average{\varrho}-\frac{1}{16\pi G}{\cal Q}_\CD - 
\frac{1}{16\pi G}\average{\CR}
\;\;\;;\;\;\;{p}^{\CD}_{\rm eff} =  -\frac{1}{16\pi G}{\cal Q}_\CD + \frac{1}{48\pi
G}\average{\CR}\;\;.
\end{equation}
\begin{equation}
\label{effectivefriedmann}
\fl
3\frac{{\ddot a}_\CD}{a_\CD} = \Lambda - 4\pi G (\varrho^{\CD}_{\rm eff}
+3{p}^{\CD}_{\rm eff})\;\;;\;\;
3H_\CD^2 =\Lambda + 8\pi G \varrho^{\CD}_{\rm eff}\;\;;\;\;
{\dot\varrho}^{\CD}_{\rm eff} + 
3H_\CD \left(\varrho^{\CD}_{\rm eff}
+{p}^{\CD}_{\rm eff} \right)=0\;.
\end{equation}
Eqs.~(\ref{effectivefriedmann}) correspond to the equations
(\ref{averageraychaudhuri}),
(\ref{averagehamilton}) and (\ref{integrability}), respectively. 

Given an equation of state of the form 
$p^\CD_{\rm eff} = \beta (\varrho^\CD_{\rm eff}, a_{\cal D})$
that relates the effective sources (\ref{equationofstate}) with a possible explicit
dependence on the volume scale factor, 
the effective Friedmann equations (\ref{effectivefriedmann})
can be solved (one of the equations (\ref{effectivefriedmann}) is redundant). 
Therefore, any question posed that is related to the evolution of scalar
characteristics
of inhomogeneous universe models may be `reduced' to finding the {\it cosmic state}
on a given spatial scale. Although formally  similar to the situation in Friedmannian cosmology, 
here the equation of state depends on the details of the evolution of inhomogeneities. 
In general it describes non--equilibrium states.

We finally wish to emphasize that these equations are limited to {\it regular} solutions: as in the non--averaged
case, the matter model `dust' generically leads to shell--crossing singularities. In the averaged
equations this fact would
be mirrored in a break of the boundary of the averaging domain or a merging of
two boundaries (Legendrian singularities), thus inducing a jump of the Euler--characteristic
of the boundary. This would especially happen for small collapsing domains and is related
to the fragmentation and merging of structures.
Here, by assumption, the domain $\CD$ must remain simply--connected.
It is possible to cure this small--scale problem by generalizing the matter model; for example, if 
multi--streaming is accounted for, an extra term related to velocity dispersion would add
up to the kinematical backreaction term. Such a term can regularize singularities as was
discussed in detail within the Newtonian framework in \cite{adhesive}.

\subsection{Derived quantities}

For later convenience we introduce a set of dimensionless average characteristics in terms
of which we shall express the solutions:
\begin{equation}
\label{omega}
\fl\qquad
\Omega_m^{\CD} : = \frac{8\pi G}{3 H_{\CD}^2} \langle\varrho\rangle_{\cal D}  \;\;;\;\;
\Omega_{\Lambda}^{\CD} := \frac{\Lambda}{3 H_{\CD}^2 }\;\;;\;\;
\Omega_{\cal R}^{\CD} := - \frac{\average{\cal R}}{6 H_{\CD}^2 }\;\;;\;\;
\Omega_{\cal Q}^{\CD} := - \frac{{\cal Q}_{\CD}}{6 H_{\CD}^2 } \;\;.
\end{equation}
We shall, henceforth, call these characteristics `parameters', but the reader should keep in 
mind that these are  functionals on $\CD$.
Expressed through these parameters the averaged Hamiltonian constraint 
(\ref{averagehamilton}) assumes the form of a `cosmic quartet':
\begin{equation}
\label{hamiltonomega}
\Omega_m^{\CD}\;+\;\Omega_{\Lambda}^{\CD}\;+\;\Omega_{\cal R}^{\CD}\;+\;
\Omega_{\cal Q}^{\CD}\;=\;1\;\;.
\end{equation}
In this set, the averaged scalar curvature parameter and the kinematical backreaction parameter
are directly expressed through $\average{\cal R}$ and ${\cal Q}_{\CD}$, respectively.
In order to compare this pair of parameters with the `Friedmannian curvature parameter' 
that is employed to interprete observational data, we can alternatively introduce the pair
\begin{equation}
\label{omeganewton}
\Omega_{k}^{\CD} := - \frac{k_{\initial\CD}}{a_\CD^2 H_{\CD}^2 }\;\;;\;\;
\Omega_{{\cal Q}N}^{\CD} := \frac{1}{3 a_\CD^2 H_\CD^2}
\int_{\initial{t}}^t \rmd t'\ {\cal Q}_\CD\frac{\rmd }{\rmd t'} a^2_\CD(t')\;\;,
\end{equation}
being related to the previous parameters by
$\;\Omega_{k}^{\CD} +\Omega_{{\cal Q}N}^{\CD}\;=\; 
\Omega_{\cal R}^{\CD} + \Omega_{\cal Q}^{\CD}$.

These parameters arise from inserting the first integral of Eq.~(\ref{integrability}),
\begin{equation}
\label{integrabilityintegral}
\frac{k_{\initial\CD}}{a_\CD^2} - \frac{1}{3 a_\CD^2} \int_{t_i}^t \,dt' \;
{\cal Q}_\CD\; \frac{d}{dt'} a^2_\CD(t')
= \frac{1}{6}\left(\,\average{\CR} + {\cal Q}_\CD\,\right) \;\;,
\end{equation}
into (\ref{averagehamilton}):
\begin{equation}
\label{averagefriedmann}
\frac{\dot{a}_\CD^2 + k_{\initial\CD}}{a_\CD^2 } - \frac{8\pi G \average{\varrho}}{3}
- \frac{\Lambda}{3} = \frac{1}{3 a_\CD^2} \int_{\initial{t}}^t \rmd t'\ {\cal Q}_\CD
\frac{\rmd }{\rmd t'} a^2_\CD(t')\;\;.
\end{equation}
This equation is formally equivalent to its Newtonian counterpart \cite{buchertehlers}, \cite{bks}.
It shows that, by eliminating the averaged curvature, the whole history of the averaged 
kinematical fluctuations acts as a source of a generalized Friedmann equation.

Like the volume scale factor $a_\CD$ and the volume Hubble functional $H_\CD$, we may 
introduce `parameters' for higher derivatives of the volume scale factor, e.g. 
the {\it volume deceleration functional} 
\begin{equation}
\label{deceleration}
q^\CD := -\frac{{\ddot a}_\CD}{a_\CD}\frac{1}{H_\CD^2} = \frac{1}{2}
\Omega_m^{\CD} + 2 \Omega_{\cal Q}^{\CD} - \Omega_{\Lambda}^{\CD}\;\;,
\end{equation}
or {\it (volume)~state finders} \cite{sahnietal,alametal} (see also \cite{evans} and  
references therein).

In this paper we shall denote all the parameters evaluated at the initial time by
the index $\initial\CD$, and at the present time by the index $\now\CD$.

\noindent
More details concerning these equations and their solutions 
may be found in \cite{buchert:grgdust,buchert:grgfluid,buchert:jgrg,buchert:static}.

\section{The morphon field}

In the above--introduced framework we distinguish the averaged matter source and averaged 
sources due to geometrical inhomogeneities stemming from extrinsic and intrinsic curvature
(backreaction terms). The averaged equations can be written as standard 
Friedmann equations that are sourced by both. Thus,
we have the choice to consider the averaged model as 
a cosmology with matter source  `morphed' by a mean field that is generated by 
backreaction terms.
We shall demonstrate below that this introduction of a `morphon field' provides a
natural description. We say `natural', because the form of the effective sources in
Eq.~(\ref{equationofstate}) shows that, for vanishing averaged scalar curvature,
backreaction obeys a stiff equation of state as suggested by the fluid analogy with a
free scalar field \cite{madsen}, \cite{buchert:grgfluid}. 
Moreover,  if we also model the averaged curvature by an effective scalar
field potential, we find that the integrability condition (\ref{integrability})
provides the evolution equation for the scalar field in this potential, and it is identical to
the Klein--Gordon equation, as we  explain now.

\subsection{Setting--up the correspondence}

We now propose to model 
the effective sources arising from geometrical degrees of freedom due to backreaction
(i.e., roughly the `vacuum' part of the effective equations)
by a scalar field $\Phi_\CD$ evolving in an 
effective potential $U_\CD := U (\Phi_\CD)$\footnote{We choose the letter $U$ for the potential
to avoid confusion with the volume functional.}, 
both  domain--dependent, as follows (recall that we have no matter pressure source
here):
\begin{equation}
\label{morphon:sources}
\varrho^\CD_{\rm eff} =: \langle\varrho\rangle_{\cal D} +
\varrho^\CD_{\Phi}\;\;\;;\;\;\;
p^\CD_{\rm eff} =: p^\CD_{\Phi}\;\;,
\end{equation}
with \cite{madsen}, \cite{mukhanov}
\begin{equation}
\label{morphon:field}
\varrho^\CD_{\Phi}=\epsilon \frac{1}{2}{{\dot\Phi}_\CD}^2 + U_\CD\;\;\;;\;\;\;p^\CD_{\Phi} =
\epsilon \frac{1}{2}{{\dot\Phi}_\CD}^2 - U_\CD\;\;,
\end{equation}
where $\epsilon=+1$ for a standard scalar field (with positive kinetic energy), and 
$\epsilon=-1$ for a phantom scalar field (with negative kinetic energy).
Thus, in view of Eq.~(\ref{equationofstate}), we obtain the following correspondence:
\begin{equation}
\label{correspondence1}
-\frac{1}{8\pi G}{\cal Q}_\CD \;=\; \epsilon {\dot\Phi}^2_\CD - U_\CD\;\;\;;\;\;\;
-\frac{1}{8\pi G}\average{\CR}= 3 U_\CD\;\;.
\end{equation} 

We see that the averaged scalar curvature directly represents the potential, whereas the
kinematical backreaction term represents `kinetic energy density' directly, 
if the averaged scalar curvature vanishes. This representation of ${\cal Q}_\CD$ is 
physically sensible, since
it expresses a balance between `kinetic energy' $E_{\rm kin}^\CD := 1/2 {\dot\Phi}^2_\CD V_\CD$
and `potential energy' $E_{\rm pot}^\CD := -U_\CD V_\CD$.
For ${\cal Q}_\CD =0$ we obtain the 
`virial condition'\footnote{For negative potential energy (positive curvature) the sign 
$\epsilon = +1$, and for positive potentials the sign $\epsilon = -1$ (phantom energy) is suggested
from the scalar virial theorem $2E_{\rm kin}^\CD = -E_{\rm pot}^\CD$.} 
$2\epsilon E_{\rm kin}^\CD + E_{\rm pot}^\CD =0$,
and so kinematical backreaction is identified as causing deviations from `equilibrium' (defined
through this balance\footnote{An alternative definition of `out--of--equilibrium' states 
uses an information theoretical measure as proposed in \cite{hosoya:infoentropy} and discussed
in the present context in \cite{buchert:static}.}). Note that Friedmann--Lema\^\i tre cosmologies
correspond to the vanishing of  ${\cal Q}_\CD$ (established `virial equilibrium'
or vanishing relative information entropy according to \cite{hosoya:infoentropy})
{\it on all scales}. The scale--dependent formulation allows to identify states in
`virial equilibrium' on particular spatial scales.
Below we learn that this `virial balance' can be stable {\it or} unstable.

Inserting (\ref{correspondence1}) into the integrability condition (\ref{integrability}) 
then implies that $\Phi_\CD$, for 
${\dot\Phi}_\CD \ne 0$, obeys the Klein--Gordon equation:
\begin{equation}
\label{kleingordon}
{\ddot\Phi}_\CD + 3 H_{\cal D}{\dot\Phi}_\CD + 
\epsilon\frac{\partial}{\partial \Phi_\CD}U(\Phi_\CD)\;=\;0\;\;.
\end{equation} 
With this correspondence the backreaction effect is formally equivalent to the dynamics of a
homogeneous, minimally coupled scalar field.
Given this correspondence we can try to reconstruct the potential in which 
the morphon field evolves.
Note that there are two equations of state in this approach, one for the morphon, 
$w^\CD_{\Phi}: = p^\CD_{\Phi}/\varrho^\CD_{\Phi}$, and the total `cosmic equation of state' 
including the matter source term, $w^\CD_{\rm eff}:= p^\CD_{\rm eff}/\varrho^\CD_{\rm eff}$.

It is also worth noting that a usual scalar field source in a Friedmannian model, attributed 
e.g. to phantom quintessence that leads to acceleration, 
will violate the {\it strong energy condition} $\varrho + 3p >0$, i.e.:
\begin{equation}
\label{energyconditionF}
3\frac{\ddot a}{a} = -4\pi G (\varrho + 3p) = -4\pi G 
(\varrho_H + \varrho_{\Phi} + 3 p_{\Phi}) \;> 0\;\;,\qquad
\end{equation}
and actually also the {\it weak energy condition} $\varrho + p >0$,
while for a morphon field both are not violated for the true content of the
Universe, that is ordinary dust matter. 
It is interesting that we can write a `strong energy condition' for the effective sources, i.e.:
\begin{equation}
\label{energyconditionQ}
\fl
3\frac{{\ddot a}_\CD}{a_\CD} =  - 4\pi G (\varrho^{\CD}_{\rm eff}
+3{p}^{\CD}_{\rm eff}) =- 4\pi G\left(\, \langle\varrho\rangle_\CD + 
\varrho_{\Phi}^\CD + 3 p_{\Phi}^\CD\,\right)
= -4\pi G \langle\varrho\rangle_\CD + {\cal Q}_\CD \;<\;0\,.
\end{equation}
While we do not need `exotic matter', the above condition will be `violated' in order to 
have volume acceleration,
${\cal Q}_\CD > 4\pi G\langle\varrho\rangle_\CD$  
\cite{kolbetal}, \cite{buchert:static}, {\it cf.} 
Subsection~\ref{subsubsect:accelerationconditions}.

\subsection{Newtonian limit}

In the Newtonian limit \cite{buchertehlers}, the above correspondence
persists. The  sources of the morphon in the effective Friedmann equations (\ref{effectivefriedmann})
are then identified as follows, {\it cf.} Eqs.~(\ref{omeganewton}) and 
(\ref{integrabilityintegral}):
\begin{eqnarray}
\label{sourcesnewton}
\varrho^\CD_{\Phi N} := \frac{1}{8\pi G}\left[\,\frac{1}{a_\CD^2} 
\int_{\initial{t}}^t \rmd t'\ {\cal Q}_\CD\frac{\rmd }{\rmd t'} a^2_\CD(t')
 -\frac{3k_{\initial\CD}}{a_\CD^2}\,\right]\;\;;\;\;\nonumber\\
p^\CD_{\Phi N} := - \frac{1}{24\pi G}\left[\,\frac{1}{a_\CD^2} 
\int_{\initial{t}}^t \rmd t'\ {\cal Q}_\CD \frac{\rmd }{\rmd t'} a^2_\CD(t') 
-\frac{3 k_{\initial\CD}}{a_\CD^2}+2{\cal Q}_\CD\,\right]\;\;.
\end{eqnarray}
However, Newtonian cosmologies suppress the morphon degrees of freedom on some fixed
large scale where the kinematical backreaction term has to 
vanish identically \cite{buchertehlers}. 
In particular, this remark applies to cosmological N--body simulations:
by construction, these simulations enforce 
`virial equilibrium' of the morphon energies on the scale of the simulation box.

The Newtonian framework also offers a concise explanation of our choice
`morphon field': the kinematical backreaction term ${\cal Q}_\CD$ can be entirely
expressed through Minkowski Functionals \cite{meckeetal} 
of the boundary of the averaging domain  \cite{buchert:jgrg}. These functionals form
a complete basis in the space of (Minkowski--)additive measures for the morphometry
of spatial sets.

\subsection{Motivation: the morphon modeling a cosmological constant} 

Let us  give a simple example.
As shown in \cite{buchert:static}, Subsect.~3.2 (also advanced as a motivation case in 
\cite{kolbetal}), the effective source ${\cal Q}_\CD$ may act
as a cosmological constant. The general condition for the corresponding exact solution of the 
effective Friedmann equations (\ref{effectivefriedmann}) 
(with $\langle\varrho\rangle_{\cal D}=0$ and $\Lambda = 0$) 
is $\average{\CR}= 6 k_{\initial\CD} /a_\CD^2 -3{\cal Q}_\CD$; ${\cal Q}_\CD = 
{\cal Q}_{\initial\CD}$; e.g., for $k_{\initial\CD} = 0$ we have:
\begin{equation}
\label{inflaton1}
{\dot\Phi}_\CD = 0\;\;\;;\;\;\; \varrho^\CD_{\Phi}= U_\CD\;\;\;;\;\;\;p^\CD_{\Phi} = -U_\CD
\;\;\;;\;\;\;U_\CD = U_{\initial\CD}  \;\;.
\end{equation}
The kinematical backreaction term assumes a constant value ${\cal Q}_{\CD}
= 8\pi G U_{\initial\CD}$, and the morphon potential mimics a (scale--dependent) cosmological constant. 
Note that the averaged curvature is non--zero, $\average{\CR}= -24\pi G U_{\initial\CD}$, 
so that simultaneously the morphon unavoidably installs a non--zero averaged scalar
curvature.

We shall now exemplify this correspondence for a new class of scaling solutions 
that contains this and other known subcases.

\section{A family of scaling solutions of the backreaction problem}

In the following, we shall study a class of solutions that prescribes backreaction and averaged 
curvature functionals in the form of scaling laws of the volume scale factor $a_{\CD}$.
If not explicitly stated otherwise, we restrict attention to the case $\Lambda = 0$ throughout,
and treat the cosmological constant as a particular morphon.

\subsection{Exact scaling solutions}

In this subsection we shall present a systematic classification of scaling behaviors 
for the cosmological models introduced previously. The averaged dust matter density 
$\langle\varrho\rangle_\CD$ evolves, for a restmass preserving domain $\CD$, as 
$\langle\varrho\rangle_\CD=\langle\varrho\rangle_{\initial\CD}\; a_\CD^{-3}$. 
Let us suppose that the backreaction term and the averaged curvature also obey scaling laws, that is:
\begin{equation}
\label{prescription}
{\cal Q}_\CD={\cal Q}_{\initial \CD}\; a_\CD^{n}\;\;\;\;;\;\;\;\;
\average{\CR}={\cal R}_{\initial \CD}\; a_{\CD}^{p}\;\;,
\end{equation}
where ${\cal Q}_{\initial\CD}$ and ${\cal R}_{\initial\CD}$ 
denote the initial values of ${\cal Q}_\CD$ and $\average{\CR}$, respectively.

Rewriting the integrability condition (\ref{integrability}),
\begin{equation}
a_\CD^{-6}\partial_{t}\left(a_\CD^{6}{\cal Q}_\CD\right)+
a_\CD^{-2}\partial_{t}\left(a_\CD^{2}\average{\CR}\right)\,=\,0\;\;,
\end{equation}
a first scaling solution of that equation is obviously provided by (\cite{buchert:grgdust}
Appendix B):
\begin{equation}
\label{ndiffpsolution}
{\cal Q}_\CD={\cal Q}_{\initial\CD }\; a_\CD^{-6}\;\;\;\;;\;\;\;\;
\average{\CR}={\cal R}_{\initial\CD}\; a_\CD^{-2}\;\;.
\end{equation} 

\vspace{-10pt}

\begin{figure}[htbp]
\begin{center}
\includegraphics[width=10cm]{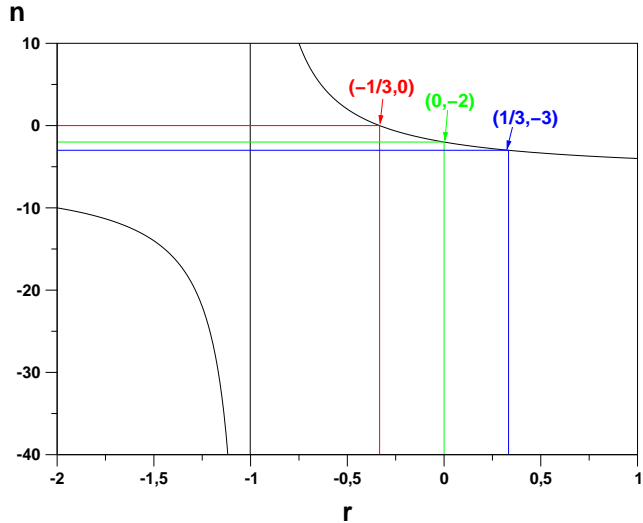}
\caption{\label{fig:r-n}
Scaling law $n$ for the backreaction as a function of the parameter $r$.}
\end{center}
\end{figure}

Moreover, this is clearly the only solution with $n\neq p$.
In the case $n=p$, we define a new `backreaction parameter' $r^\CD$ (that can be chosen
differently for a chosen domain of averaging\footnote{For notational ease we henceforth
drop the index $\CD$ and simply write $r$.}) such that ${\cal Q}_{\initial\CD}\propto
{\cal R}_{\initial\CD}$; the solution reads: 
\begin{equation}
\label{backreactionparameter}
{\cal Q}_\CD \;=\;r \;\average{\CR}\;=\;r \;{\cal R}_{\initial\CD} \,a_\CD^{n}\;\;, 
\end{equation}
where ({\it cf.} Fig.~\ref{fig:r-n})
\begin{equation}
\label{neqpsolution}
n\;=\;-2 \frac{(1+3r)}{(1+r)}\;\;,
\end{equation}
with $r\neq-1$. The case $r=-1$ is not represented as a solution in this class; this line
of states degenerates to a point corresponding to a model with Einstein--de Sitter 
kinematics, i.e. it has
vanishing backreaction and vanishing averaged curvature.
Note here, that the vanishing of ${\cal Q}_\CD$, if required on all domains, is necessary, 
but also sufficient for the reduction of the averaged equations to Friedmann--Lema\^\i tre
cosmologies. This limiting case also appears in the relations among the cosmological
parameters, discussed after the next subsection.

\subsection{Discussion of the scaling solutions}
 
The solution (\ref{ndiffpsolution}) corresponds to the case where the backreaction and the 
averaged scalar curvature evolve independently, leading to an averaged
curvature similar to a constant `Friedmannian curvature' and an additional
term that scales as $\langle\varrho\rangle_{\cal D}^2$. 
As the universe model expands, this solution is
rapidly equivalent to a pure constant curvature term, we may say that this solution
represents (and maintains) a near--Friedmannian state (${\cal Q}_\CD$ decays much more rapidly
$\propto V_\CD^{-2}$ compared with the averaged density $\propto V_\CD^{-1}$).
Therefore, we shall not model  this solution by a scalar field when describing the late--time 
dust--dominated Universe. 
It is interesting to note that this solution exhibits
the same late--time behavior as the long--wavelength part of the solution
found through the gradient expansion approximation scheme in \cite{kolbetal}.

On the contrary, the solutions (\ref{backreactionparameter}) entail a strong coupling between 
kinematical backreaction and averaged scalar curvature.
The coupling itself must be considered a generic property; it has been
identified as being responsible for a much slower decay of kinematical fluctuations in an
expanding universe model on the cost of averaged scalar curvature. 
This is a genuinly relativistic property.
It was argued in
\cite{buchert:darkenergy} that it is this possibility which is needed for an
explanation of Dark Energy through backreaction today, provided the initial conditions are 
appropriate.

The scaling solutions (\ref{backreactionparameter}) can be employed to represent
generic features of backreaction-- or curvature--dominated cosmologies
(while density fluctuations must not be large). Such cosmologies may 
significantly deviate from a standard Friedmann cosmology with regard to the temporal 
evolution of their parameters. We note that, even if deviations
in the volume scale factor $a_\CD$ at a final time may not be large, deviations in
in its history, i.e. its time--derivatives,
in particular its second time--derivative related to the cosmological parameters, may be large
(this insight is a result of a detailed analysis in Newtonian cosmology \cite{bks}). 

Let us  make a general remark concerning the solution subspace of the scaling solutions given above.
We appreciate that the polynomial nature of (\ref{prescription}),
together with the form of the integrability condition (\ref{integrability}), implies 
that any linear combination of these solutions provides a new solution. 
In particular, one can always add a constant curvature term to a particular solution.
We also infer that only the case $r=0$ ($n=-2$)  implies a (scale--dependent) `Friedmannian' 
evolution of the (physical) curvature parameter $\Omega^\CD_{\cal R}$. 
Only the singular case $r=-1$ would imply a vanishing
`Friedmannian' curvature parameter (see below). If we would require $r=-1$ on all scales, then the model
reduces to the Euclidean case
(everywhere vanishing 3--Ricci curvature).

We resume this discussion in the next section with the help of concrete examples.
There we also provide an illustration and further discussion of the subspace 
defined by the scaling solutions.

Finally let us note that, 
while the scaling solutions and scenarios investigated here and below satisfy the 
averaged equations of motions, these models are still phenomenological and not 
derived non--perturbatively from the fundamental theory, i.e. 
there is no guarantee that corresponding realistic solutions of 
the original inhomogeneous Einstein equations could be found that satisfy the 
assumed scaling laws.

\subsection{Some relations among cosmological parameters}

We here write some useful relations among the dimensionless cosmological
parameters, as they were introduced earlier, Eqs.~(\ref{omega}) and (\ref{omeganewton}). 
For the scaling solutions (\ref{neqpsolution}) we have:
\begin{equation}
\label{omegascaling1}
\Omega_{\cal Q}^{\CD} \;=\; r\; \Omega_{\cal R}^{\CD} \;\;,
\end{equation}
and for the `Friedmannian curvature parameter' we find (by integration of (\ref{integrabilityintegral})
for the solutions (\ref{neqpsolution})):
\begin{eqnarray}
\label{omegascaling2}
\Omega_{k}^{\CD} =  \Omega_{\cal R}^{\CD} + \Omega_{\cal Q}^{\CD}
-\frac{1}{3 a_\CD^2 H_\CD^2}
\int_{\initial{t}}^t \rmd t'\ {\cal Q}_\CD\frac{\rmd }{\rmd t'} a^2_\CD(t') \;=\nonumber\\ 
(1+r)\,a_\CD^{\frac{4r}{(1+r)}}\,\Omega_{\cal R}^{\CD}\,=\,
(1+r)\,\frac{H_{\initial\CD}^2}{H_\CD^2}\,a_\CD^{-2}\,\Omega_{\cal R}^{\initial\CD}\;\,.
\end{eqnarray}
(The latter equation follows by noting that $\Omega_{\cal R}^{\CD}=
\Omega_{\cal R}^{\initial\CD}\,(H_{\initial\CD}^2 / H_\CD^2 )\, a_\CD^n$ and
$4r/(1+r) = -(n+2)$).
Eq.~(\ref{omegascaling2}) now explicitly shows that the generally held view that 
$\Omega_{k}^{\CD}$ models the averaged curvature is a misperception: as soon as 
kinematical backreaction is relevant (itself or its time--history), the averaged curvature may
evolve very differently compared with a constant--curvature model.
Evaluating (\ref{omegascaling2}) at initial time, we find that initial data differ only by the 
parameter $r$ in this class of solutions, which eases their observational determination:
\begin{equation}
\label{omegascaling0}
\Omega_{k}^{\initial\CD} = (1+r)\Omega_{\cal R}^{\initial\CD} \;\,,\;{\rm i.e.}\;,\;\;
k_{\initial\CD} = \frac{(1+r)}{6}{\cal R}_{\initial\CD}\;\;.
\end{equation}
Since in the sum $\Omega_{\cal R}^{\CD} + \Omega_{\cal Q}^{\CD} =
\Omega_{{\cal Q}N}^{\CD} + \Omega_{k}^{\CD}$, the `Newtonian' parameter 
$\Omega_{{\cal Q}N}^{\CD}$, {\it cf.} Eq.~(\ref{omeganewton}), 
would be interpreted as a cosmological constant parameter, $\Omega^F_{\Lambda}$, 
in a `Friedmannian fitting model', we can
already from these relations infer that with $\Omega^\CD_k = -k_{\initial\CD}/
(H_\CD^2 a_\CD^2 )$,
$\Omega^{\initial\CD}_k = (1+r) \Omega^{\initial\CD}_{\cal R}$, and  
$\Omega^\CD_k = \Omega^{\initial\CD}_k \, (H_{\initial\CD}^2 / H_\CD^2 a_\CD^2)$,
the value of a fitted $\Lambda-$parameter would directly depend on the initial datum for
$ \Omega^{\CD}_k$ according to the relation
\begin{equation}
\label{constraininglambda}
\Omega_{\cal R}^{\CD} + \Omega_{\cal Q}^{\CD} \;=\; 
(1+r)\frac{\Omega_{\cal R}^{\initial\CD}H_{\initial\CD}^2}{H_\CD^2}\,a_\CD^n \;=\;
\Omega^{\initial\CD}_k \frac{H_{\initial\CD}^2}{H_\CD^2}\,a_\CD^2 \;=\;
 \Omega^{\CD}_k \,a_\CD^{n+2}\;\;,
\end{equation}
so that today 
\begin{equation}
\label{constrainttoday}
\Omega_{\cal R}^{\now\CD} + \Omega_{\cal Q}^{\now\CD} 
\;=\; \Omega^F_{\Lambda} + \Omega^{\now\CD}_k \;=\;
 \Omega^{\now\CD}_k \,a_{\now\CD}^{n+2}\;\;.
\end{equation}
With this relation we can determine the dependence of the scaling
solution parameter $r$ on the cosmological parameters today, expressed through
the `Friedmannian fitting parameters' $\Omega^F_{\Lambda}$ and 
$\Omega^{\now\CD}_k$. This allows us to 
put constraints on $r$, which will be done in 
Subsections~\ref{subsect:observationalconstraints} and
\ref{subsect:concrete}. 

We see that in these estimates we also need the correct evolution of $a_\CD$ in order
to calculate $a_{\now\CD}$. This is possible for the particular class of scaling solutions
we consider (we have numerically evaluated the evolution for the volume scale factor
below); in general it requires a detailed model for the evolution of inhomogeneities. 

Finally, we emphasize that the above relations are a result of our `single--scaling' model ansatz.
More flexible models arise by superimposing scaling solutions (and so modeling
the dynamics more realistically). The relation (\ref{constraininglambda}) is a consequence of
this: in general, initial data for the 3--Ricci curvature and the integration constant $k_{\initial\CD}$
can be independently chosen. The reader should therefore make up their mind 
about a particular choice of superposition of scaling solutions 
that they would like to implement. 

In this line our model is maximally conservative concerning the amount of 
`early' Dark Energy \cite{doranetal}, 
i.e. the value of $\Omega^\CD_{{\cal Q}N}$, interpreted as a (constant) $\Lambda-$ parameter at the
present time evolves as a time--dependent `cosmological term'. 
Its initial value can be calculated:
\begin{equation}
\label{earlydarkenergy} 
\Omega^{\initial\CD}_{{\cal Q}N} = \Omega_{\cal R}^{\initial\CD} + 
\Omega_{\cal Q}^{\initial\CD}-\Omega^{\initial\CD}_k \;=\;  (1+r) \Omega_{\cal R}^{\initial\CD}
-  \Omega^{\initial\CD}_k \;=\;0\;\;. 
\end{equation}
Thus, we implicitly require the initial contribution of this term
to vanish, while actually a value in the range of a few percent would be allowed by observational
constraints \cite{caldwell&mukhanov}, \cite{caldwell&wetterich}, \cite{caldwell&linder}.

\section{Morphon--quintessence}

The obvious candidates for scalar field models, which come into the fore in the 
situation of a dust--dominated Universe, are quintessence models.
We shall now exploit the proposed correspondence to explicitly reconstruct the
scalar field dynamics. 
Since quintessence models aim at mimicking a repulsive component in the cosmological
evolution, it is natural that a `working model' of quintessence would, via the proposed 
correspondence of a `morphon--quintessence', also lead to a `working model' of 
backreaction.
For the purpose of concretizing the correspondence,  
let us now consider solutions of the type (\ref{backreactionparameter}).
We will show that they can be put in correspondence with a one--parameter
family of homogeneous scalar field solutions that act as
standard or phantom quintessence fields.  

\subsection{Reconstructing the potential of the morphon field}

The scaling solutions (\ref{backreactionparameter}) provide:
\begin{equation}
\label{identification1}
\fl
{\dot\Phi}^2_\CD=-\epsilon\frac{{\cal R}_{\initial\CD}}{8\pi
  G}\left(r+\frac{1}{3}\right)\;a_\CD^{n}\;\;;\;\;
U_\CD=-\frac{{\cal R}_{\initial\CD}}{24\pi G}\;a_\CD^{n}\;\;;\;\; 
n=-2\frac{(1+3r)}{(1+r)}\;;\;{\cal Q}_{\initial\CD} = r
{\cal R}_{\initial\CD}\;.
\end{equation}
This correspondence defines a scalar field evolving in a {\it positive} potential, if ${\cal R}_{\initial\CD}<0$ (and in a {\it negative} potential if ${\cal R}_{\initial\CD}>0$), 
and a {\it real} scalar field, if $\epsilon{\cal R}_{\initial\CD}(r+1/3)<0$. 
In other words, if ${\cal R}_{\initial\CD}<0$ we have a priori a phantom field
for $r<-1/3$ and a standard scalar field for $r>-1/3$; if  ${\cal
  R}_{\initial\CD}>0$, we have a standard scalar field for $r<-1/3$ and a
phantom field for  $r>-1/3$.

The system (\ref{identification1}) can be inverted to reconstruct the
potential of the morphon field. 
Indeed, the kinetic term of the scalar field can be expressed in terms of
$d\Phi_\CD/da_\CD$, and this equation can be explicitly integrated, for
$\langle\varrho\rangle_{\initial\CD}\neq 0$, leading to:
\begin{eqnarray}
\label{phi/a}
\Phi_\CD(a_\CD)=\frac{2\sqrt{\epsilon(1+3r)(1+r)}}{(1-3r)\sqrt{\pi G}}\;
{\rm arsinh}\left(\sqrt{\frac{-(1+r){\cal R}_{\initial\CD}}{16\pi G\langle\varrho\rangle_{\initial\CD}}
a_\CD^{\frac{(1-3r)}{(1+r)}}}\right)\nonumber\\ \qquad\quad\;\,\,=\,
\frac{\sqrt{-2\epsilon n}}{(n+3)\sqrt{\pi G}}\;{\rm arsinh}\left(\sqrt{(1+r)\gamma_{\CR m}^\CD}\right)\;\;,
\end{eqnarray}
where we have defined the fraction $\gamma_{\CR m}^\CD$ of the curvature and density parameters:
\begin{equation}
\label{gammaR}
\gamma_{\CR m}^\CD := \frac{\Omega_{\cal R}^\CD}{\Omega_m^\CD} = 
\gamma_{\CR m}^{\initial\CD} \,
a_\CD^{(n+3)}\;\;\;;\;\;\;(n+3) = \frac{(1-3r)}{(1+r)}\;\;;\;\;
\gamma_{\CR m}^{\initial\CD} := \frac{\Omega_{\cal R}^{\initial\CD}}{\Omega_m^{\initial\CD}}\;\;.
\end{equation}
In this relation, necessarily, $r\neq-1/3$ and $r\neq-1$. We immediately find that $\Phi_\CD(a_\CD)$ is an increasing
function of $a_\CD$.
Then, inverting that relation and inserting the result into the expression for
the potential, we obtain the explicit form of the self--interaction term of the
scalar field: 
\begin{eqnarray}
\label{pot1}
\fl
U(\Phi_\CD)=\frac{-(1+r){\cal R}_{\initial\CD}}{24\pi G}\left((1+r)\gamma_{\CR
  m}^{\initial\CD}\right)^{2\frac{(1+3r)}{(1-3r)}} \,\sinh^{-4\frac{(1+3r)}{(1-3r)}}
\left(\frac{(1-3r)\sqrt{\pi G}}{\sqrt{\epsilon(1+3r)(1+r)}}\Phi_\CD\right)\nonumber\\
\fl
\qquad\quad\,=\, \frac{2(1+r)}{3}\left((1+r)\gamma_{\CR m}^{\initial\CD} \right)^{\frac{3}{(n+3)}} 
\;\langle\varrho\rangle_{\initial\CD}\,
\sinh^{\frac{2n}{(n+3)}}\left(\frac{(n+3)}{\sqrt{-\epsilon n}}\sqrt{2\pi G} \Phi_\CD \right)\;\;,
\end{eqnarray}
where $\langle\varrho\rangle_{\initial\CD}$ is the initial averaged restmass density of 
dust matter, and where the restrictions introduced above still hold, with the
new constraint $r\neq1/3$.
To sum up, in order to obtain a consistent description in terms of a {\it real--valued} scalar field, 
we must have ${\cal R}_{\initial\CD}>0$ and $\epsilon=+1$ for $r<-1$, and 
${\cal R}_{\initial\CD}<0$ for $r>-1$, with $\epsilon=-1$, if $-1<r<-1/3$ and $\epsilon=+1$, 
if $r>-1/3$.
One can immediately notice that the energy scale of the scalar field potential
is determined by the averaged matter density:
the scales that determine the scalar field dynamics are fixed by
the matter distribution.   
As a result of our correspondence to quintessence models and in view of many results 
that were obtained in this field, the above potential can also be found in \cite{sahnistarobinskii}
(their Eq.~(121) with a typo corrected in \cite{sahnietal}\footnote{Thanks to 
Varun Sahni, who has pointed this out to us.}) 
and, e.g. \cite{urenamatos}.

We  now study the cases that were excluded in the above derivations.
First, let us consider the case of vanishing matter source. 
The correspondence given by equations (\ref{phi/a}) and (\ref{pot1})
holds in the presence of a non--vanishing matter field, but one can also reconstruct the
scalar field cosmology for the vacuum. Setting $\langle\varrho\rangle_{\initial\CD}=0$ 
in the equations (\ref{averagehamilton}) and (\ref{averagefriedmann}), and applying the 
same procedure as the one described above, one finds:
\begin{equation}
\label{potvac}
U(\Phi_\CD)=-\frac{{\cal R}_{\initial\CD}}{24\pi G}
\exp\left(-4\sqrt{\frac{\epsilon (1+3r)}{(1+r)}}\sqrt{\pi G}\,\Phi_{\CD}\right)\;.
\end{equation}
Up to the renormalization factor 
$(1+r)\left((1+r)\gamma_{\CR m}^{\initial\CD}\right)^{2\frac{(1+3r)}{(1-3r)}}$ 
that reflects the presence of matter, this is exactly the solution (\ref{pot1}) in the case
$\Phi_{\CD}\rightarrow +\infty$.

As noted above, the morphon field can be interpreted as representing the
effect of the averaged geometrical degrees of freedom; then the comparison of its
potential in the presence of matter and in the vacuum tells us, how the matter
field influences the backreaction terms: it affects the energy scale
through a simple factor depending on the initial averaged matter density, and so modifies
the dynamics of the morphon when the domain volume is small (because $\Phi_{\CD}$ is
an increasing function of $a_{\CD}$; the limit $\Phi_{\CD}\rightarrow
+\infty$ corresponds to $a_{\CD}\rightarrow +\infty$.) On the contrary, when
the domain volume becomes big, the dynamics of backreaction is similar to its
dynamics in vacuum, which is natural because the averaged matter density is then 
diluted.

Second, we discuss the three cases of the backreaction parameter $r$ 
that were not considered until 
now: $r=\pm 1/3$ and $r=-1$.
In the case $r=1/3$, the solution reads
${\cal Q}_\CD=1/3\average{\CR}\propto a_\CD^{-3}$; this corresponds 
to a scale--dependent Einstein--de Sitter scenario with a renormalized initial dust density 
$\langle\varrho\rangle_{\initial\CD}-{\cal R}_{\initial\CD}/36\pi G$ 
({\it cf.} Appendix A). This model has zero effective pressure $p_{\rm eff}$.
The case $r=-1/3$ leads to ${\cal Q}_\CD=-1/3\average{\CR}=const.$, 
which is equivalent to a scale--dependent Friedmanian scenario with a cosmological constant: 
$\Lambda={\cal Q}_{\initial\CD}= -{\cal R}_{\initial\CD}/3$ (that was our motivating
example).
The case $r=-1$ corresponds to a strict compensation between the kinematical backreaction and 
the averaged scalar curvature. It leads to a scale--dependent Friedmann model 
with only a dust matter source, in other words, to a 
scale--dependent Einstein--de Sitter model. 

The above three cases appear as limiting cases of the scalar field model.

The scaling solutions correspond to specific scalar field models with a constant 
partition of energy 
between the kinetic and the potential energies of the scalar field. 
Indeed, if we define the 
kinetic energy by $E^\CD_{\rm kin}:=\frac{1}{2}{{\dot\Phi}_\CD}^2 V_\CD$ 
and the potential energy 
by $E^\CD_{\rm pot}:=-U(\Phi_\CD) V_\CD$ as before, 
we find the following `balance condition' for the scalar field 
representation of backreaction and averaged scalar curvature 
in the case of the scaling solutions:
\begin{equation}
\label{virial}
E^\CD_{\rm kin}+\frac{(1+3r)}{2\epsilon}E^\CD_{\rm pot}\;=\;0\;\;.
\end{equation}
We previously discussed the case $r=0$ (`zero backreaction') for which this 
condition agrees with the standard scalar virial theorem.
This balance between kinetic and potential energies is well known in the
context of scaling solutions of quintessence 
(see \cite{liddle98,Peebles2003} and references therein).

\vspace{5pt}

\noindent
Finally, the effective equation of state for this morphon field is constant and given by:
\begin{equation}
\label{eosmorphon1}
w^\CD_{\Phi}=-\frac{1}{3}\frac{(1-3r)}{(1+r)}\,=\,-\frac{1}{3}(n+3)\;\;,
\end{equation}  
which is less than $-1/3$, iff $r\in]-1;0[$.
The overall `cosmic equation of state' including the matter source term is given by:
\begin{eqnarray}
\label{eosmorphon1plusmatter}
\fl
\qquad w^{\cal D}_{\rm eff}:=\frac{p_{\rm eff}^{\cal D}}{\varrho_{\rm eff}^\CD}
=-\frac{1}{3}\frac{(1-3r)a_{\CD}^{\frac{1-3r}{1+r}}}{(1+r)a_{\CD}^{\frac{1-3r}{1+r}}
+1/\gamma_{\CR m}^{\initial\CD}}\,=\,w^\CD_{\Phi}\frac{1}{1 + a_\CD^{-(n+3)}/
\gamma_{k m}^{\initial\CD}}\;\;,\\
\qquad{\rm with}\;\;\;\gamma_{k m}^{\initial\CD}:= \frac{\Omega_k^{\initial\CD}}
{\Omega_m^{\initial\CD}}= (1+r)\gamma_{\CR m}^{\initial\CD}\;\;.
\label{gammaK}
\end{eqnarray}
Or, equivalently:
\begin{equation}
\label{eosmorphon1plusmatter2}
\qquad w^{\cal D}_{\rm eff}\;=\;w^\CD_{\Phi}\,(1-\Omega_m^{\CD})\;\;.
\end{equation}
Thus, the `cosmic state' asymptotically evolves, for an expanding universe model with 
$(n+3)=(1-3r)/(1+r) >0$,  into $w^{\cal D}_{\rm eff} \rightarrow -(n+3)/3 = w^\CD_{\Phi}$.
A necessary condition for the scalar field part to dominate 
the expansion of the universe model at late times is 
$n=-2 (1+3r)/(1+r)>-3$, which implies $r<1/3$. 
In that case, since the equation of state $w^\CD_{\Phi}$ is constant, 
a universe model dominated by 
backreaction and averaged curvature approaches the following evolution of the volume 
scale factor (note $a_{\initial\CD} = a_\CD (t_i) =1$):

\begin{figure}
\begin{center}
\includegraphics[width=12cm]{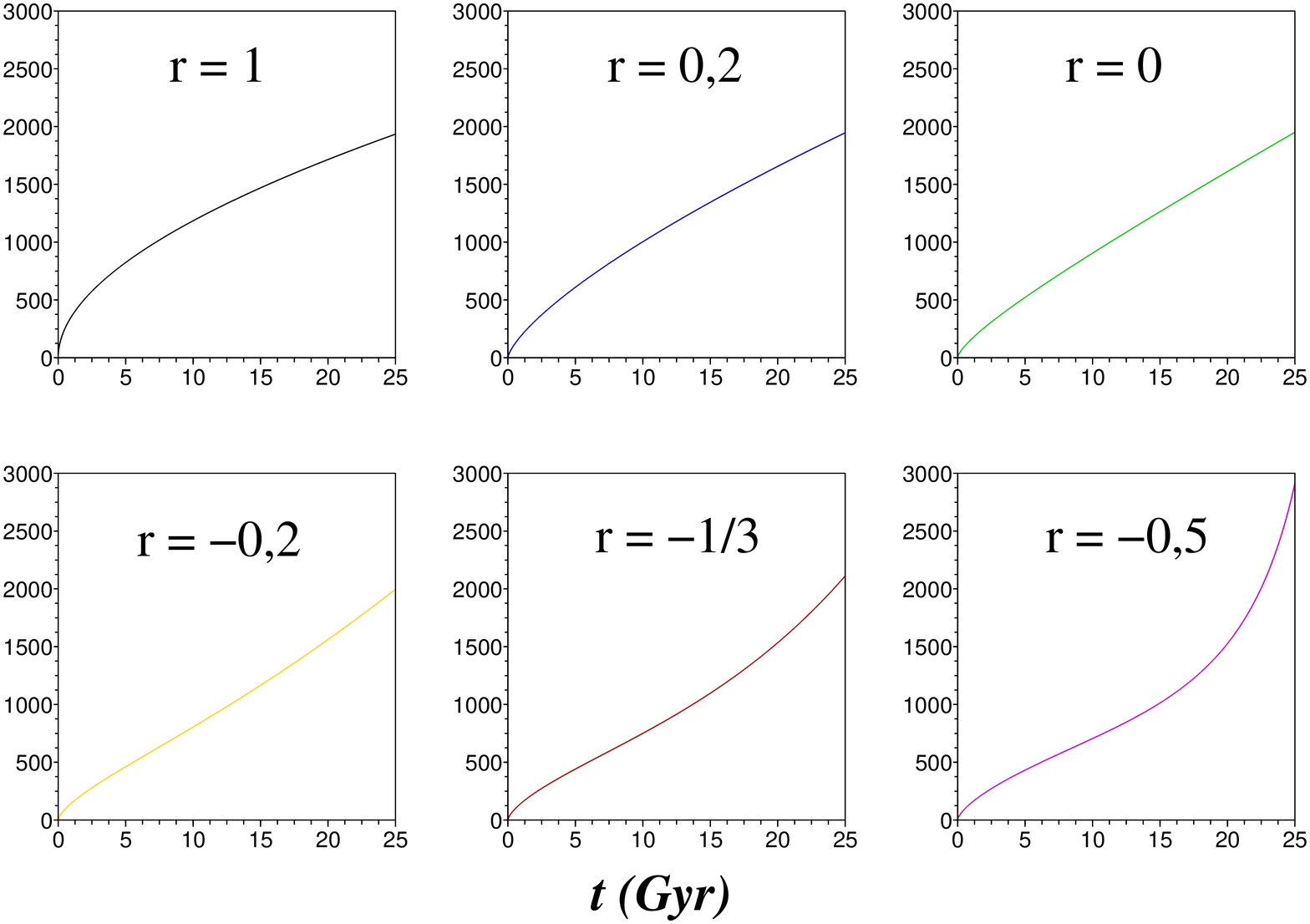}
\caption{\label{fig:scalefactor}
Volume scale factor $a_\CD$ as a function of $t$ for various scaling solutions;
note that all the models shown have vanishing cosmological constant. The
models had been integrated with $a_{\now\CD}=1000$, $H_{0}=70$ km/s/Mpc,
$\Omega_{m}^{\now\CD}=0.27$ and 
$\Omega_{{\cal Q}}^{\now \CD}+\Omega_{{\cal R}}^{\now \CD}=0.73$.
A classification scheme comprising the various cases will be provided below.
Within this scheme the first row displays examples of Case D, Case C and the Einstein--de Sitter solution;
the second row an example of Case B for a model of quintessence, 
the solution mimicking a cosmological constant,
and an example of Case A for a model of phantom quintessence.}
\end{center}
\end{figure}
 
\begin{eqnarray}
\label{asymptoticsolution}
a_{\CD}(t)\;=\; \left(1+C(t-t_i)\right)^{(1+r)/(1+3r)}\,=\,\left(1+C(t-t_i)\right)^{-2/n}
\;\;;\nonumber\\
{\rm with}\qquad
C:= \frac{4}{9}\frac{(1+3r)}{(1+r)}\sqrt{-(1+r){\cal R}_{\initial\CD}/6}\,=\,-\frac{2n}{9}
\sqrt{-k_{\initial\CD}}\;\;.
\end{eqnarray}
For the class of scaling solutions we can obtain the time--evolution of the
volume scale factor, and we appreciate that the scaling solutions explore possibilities
similar to a Friedmannian evolution {\it with} a cosmological constant.
Figure~\ref{fig:scalefactor} illustrates this for some chosen values, where the 
density parameter is held fixed for comparison with our later analysis.
For this figure we numerically integrate Eq.~(\ref{averagehamilton}) for $\Lambda = 0$
and with the scaling solutions inserted:
\begin{eqnarray}
\label{averagehamiltonscaling}
\left( \frac{{\dot a}_\CD}{a_\CD}\right)^2 \;=\; \frac{8\pi G}{3}\frac{
\langle\varrho\rangle_{\initial\CD}}{a_\CD^3}
- \frac{(1+r)}{6}\langle{\cal R}\rangle_{\initial\CD}a_{\CD}^{n} \nonumber\\
=\;\frac{8\pi G}{3}\frac{\langle\varrho\rangle_{\initial\CD}}{a_\CD^3}-
k_{\initial\CD}a_\CD^n \;=\;H_{\initial\CD}^2\,\left(
 \frac{\Omega_m^{\initial\CD}}{a_\CD^3}  +
\Omega_k^{\initial\CD}a_\CD^n\right)\;.
\end{eqnarray}
The integration is performed using a non--stiff predictor--corrector Adams
method provided by {\it Scilab}, with $a_{\now\CD}=1000$, $H_{\now\CD}=70 \mbox{
km/s/Mpc}$, $\Omega_{m}^{\now\CD}=0.27$, and $\Omega_{\cal
Q}^{\now\CD}+\Omega_{\CR}^{\now\CD}=0.73$, which yields for the constants in
the previous equation:
$8\pi
G\langle\varrho\rangle_{\initial\CD}/3=a_{\now\CD}^{3}H_{\now\CD}
\Omega_{m}^{\now\CD}$ and $(1+r){\cal R}_{\initial\CD}/6
=-a_{\now\CD}^{-n}H_{\now\CD}(\Omega_{\cal Q}^{\now\CD}+
\Omega_{\CR}^{\now\CD})$.

\subsection{Classification of the scaling solutions}

The following classification summarizes the results:
\begin{itemize}
\item {\it Case A}: for $r\in ]-1;-1/3[$ we have a phantom scalar field with a positive potential of the form
$U_\CD \propto \sinh^{\beta}(\alpha \Phi_\CD)$ with $\beta>0$; these models may lead to 
an accelerated expansion. A particular example of this type is analyzed below.
\item for $r=-1/3$ we have a model that corresponds to a scale--dependent cosmological constant 
given by $\Lambda={\cal Q}_{\initial\CD}= -1/3 {\cal R}_{\initial\CD}$.
\item for $r>-1/3$ we have a standard scalar field. Here, we can distinguish various cases:
\begin{itemize}
\item {\it Case B}: for $r\in ]-1/3;0[$, the potential is positive and of the form $\sinh^{\alpha_1}(\Phi_\CD)$, with 
$\alpha_1 \in ]-4;0[$, leading to a quintessence field that may produce an accelerated expansion.
At the beginning, when $\Phi_\CD\sim 0$, the potential is an inverse power--law
one, corresponding to the so--called Ratra--Peebles potential \cite{Ratra88}
$U(\Phi_\CD)\propto \Phi_\CD^{-\alpha_2}$, with $\alpha_2=4(1+3r)/(1-3r)$; when
$\Phi_\CD$ becomes big, the potential is equivalent to an exponential
potential, also well--known in the quintessence context \cite{Ratra88,Lucchin85}:
$U(\Phi_\CD)\propto {\rm exp} (-\alpha_3 \Phi_\CD)$, with
$\alpha_3 =2\sqrt{(1+3r)/(1+r)}$. It should be noted that the problem
emphasized in \cite{liddle98,Peebles2003} for this kind of potential (that is the
necessarily small scalar field density because of primordial nucleosynthesis constraints)
doesn't hold here, since this potential arises during the matter--dominated era as a result of
backreaction, so that in such a scenario the scalar field is significantly sourced only during
late stages of the matter--dominated era.    
The more $r$ approaches $-1/3$, the more this field mimics a cosmological constant.
\item {\it Case C}: for $r\in [0;1/3[$ the potential still behaves as $\sinh^{\alpha_4} (\Phi_\CD)$, but with 
$\alpha_4 <-4$. This potential is too stiff and the model cannot produce an accelerated expansion 
($w^\CD_{\Phi}>-1/3$).
\item for $r=1/3$, the model is equivalent to a standard Einstein--de Sitter model with a 
scale--dependent and renormalized initial dust density and zero effective pressure $p_{\rm eff}$
({\it cf.} Appendix A).
\item {\it Case D}: for $r>1/3$, the potential is of the form $\sinh^{\alpha_5}(\Phi_\CD)$ with $\alpha_5>4$ 
and the model cannot produce an accelerated expansion.
\end{itemize}
\item{\it Case E}: for $r<-1$, we have a standard scalar field rolling in a negative 
potential that is not bounded from below. Whereas such scalar fields are pathological 
when considered like fundamental scalar fields, this solution may be physical in the
backreaction context. Indeed, the four preceeding models all correspond to 
${\cal R}_{\initial\CD}<0$, and this one corresponds to 
${\cal R}_{\initial\CD}>0$. We expect that more realistic solutions, modelled e.g. by a suitable
superposition of scaling solutions, could provide potentials with minima, hosting `bound states'.
\end{itemize}

\subsection{The solution space explored by the morphon}

\begin{figure}[htbp]
\begin{center}
\includegraphics[width=14cm,height=14cm]{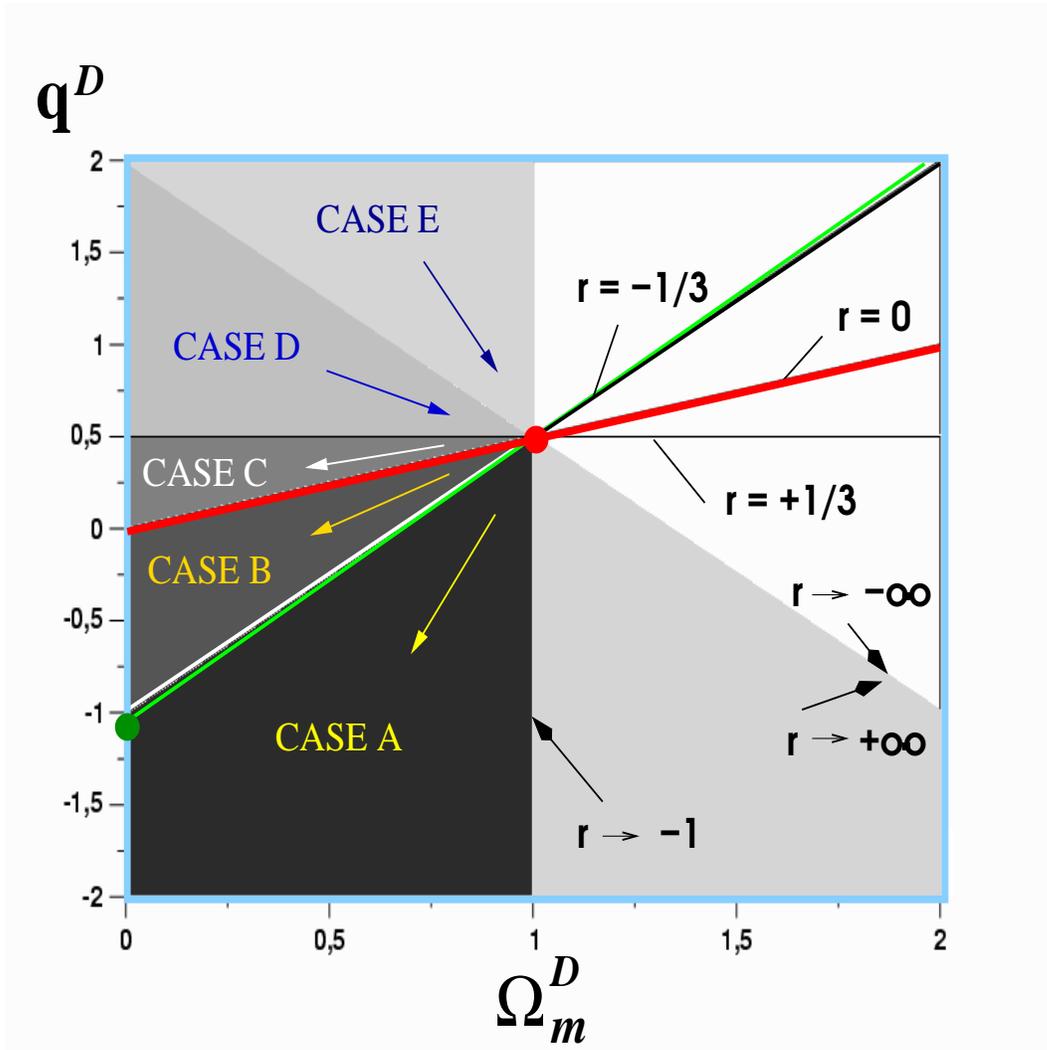}
\caption{\label{fig:cosmicstates}
This `cosmic phase diagram' 
is valid for all times and on all scales, i.e. it can be read as 
a diagram for the corresponding parameters `today' on the scale of the observable 
Universe.  All the scaling solutions are represented by straight lines passing
through the Einstein--de Sitter model in the center of the diagram ($1/2;1$). 
The vertical line corresponding to ($q^{\CD};1$) is not associated with a solution of the 
backreaction problem; it degenerates to the Einstein--de Sitter model ($1/2
;1$). This line forms a `mirror': inside the cone (Case E) there  are
solutions with $\Omega^{\CD}_m>1$ that cannot be related to any real--valued scalar field,
but are still of physical interest in the backreaction context (models with positive averaged
scalar curvature).
Models with `Friedmannian kinematics', but with renormalized parameters form the line 
$r=1/3$. The line $r=0$ are models with no backreaction 
on which the parameter $\Omega^\CD_k$ varies
(scale--dependent `Friedmannian models'). Introducing $\Lambda$ would just shift the whole
diagram down.
Below the line $r=0$ in the `quintessence phase'
we find effective models with subdominant shear fluctuations
(${\cal Q}_\CD$ positive, $\Omega^\CD_{\cal Q}$ negative).The line $r=-1/3$
mimics a `Friedmannian model' with scale--dependent cosmological constant.
The line below $r=-1/3$ in the `phantom quintessence phase'
represents the solution inferred from SNLS data ({\it cf.} 
Subsection~\ref{subsect:concrete}),
and the point at ($q^{\CD};\Omega^\CD_m) = (-1.03; 0)$ 
locates the late--time attractor associated with this solution.}
\end{center}
\end{figure}

The various cases listed above appear as `cosmic states' 
that separate `cosmic phases', illustrated in the following phase diagram, 
Fig.~\ref{fig:cosmicstates}.
To understand this diagram, we remark that the solution space for the effective
cosmologies that are sourced by dust matter and `morphed' by the scaling solutions
form one--dimensional subsets in the two--dimensional
space that is defined by Hamilton's constraint 
(taken at fixed spatial scale $\CD$ and at $\Lambda = 0$),
$\Omega^\CD_m + \Omega^\CD_{\cal Q}+\Omega^\CD_{\CR} =1$.
Also, (scale--dependent) `Friedmannian' models, characterized by the
backreaction parameter $r=0$ and $\Lambda =0$,  form a 
one--dimensional subset defined by $ \Omega^\CD_m + \Omega^\CD_k =1$.
With the scaling solutions 
we are also restricted to measure zero sets, but we have a 
one--parameter family of them which allows us to explore the  solution space.

To illustrate the solution plane, 
we plot, instead of $\Omega^\CD_{\cal Q}$ 
(related to $\Omega^\CD_{\CR}$) the volume deceleration parameter 
$q^{\CD}= 1/2 \Omega^\CD_m + 2  \Omega^\CD_{\cal Q}$,
Eq.~(\ref{deceleration}), as a function of the only free parameter $\Omega^\CD_m$
in the `cosmic phase diagram' Figure~\ref{fig:cosmicstates}.
For the scaling solutions we simply have:
\begin{equation}
\label{decelerationscaling}
q^{\CD}=\frac{2r}{1+r}+\frac{1-3r}{2(1+r)}\;\Omega^{\CD}_{m}\;=\;
-\frac{n+2}{2}+\frac{n+3}{2}\;\Omega^{\CD}_{m}\;\;.
\end{equation}

A priori, the scaling solutions can describe the whole plane, including
cosmologies with $\Omega_{m}^{\CD}>1$. Nevertheless, 
$\Omega_{m}^{\CD}>1$ implies,
from Hamilton's constraint~(\ref{hamiltonomega}) with $\Lambda=0$, that
$(1+r){\cal R}_{\initial \CD}>0$, which is exactly the opposite
condition to the one that holds in the correspondence with a real--valued scalar
field. In other words, a real--valued morphon field is only defined for
$\Omega_{m}^{\CD}<1$, and we shall concentrate on this class of models
in the following. 

Since $\Omega^{\CD}_{\cal Q} + \Omega^{\CD}_{\cal R}$ would be 
interpreted, in a Friedmannian `fitting model', as 
$\Omega^{F}_{\Lambda} + \Omega^{\now\CD}_k$, with negligible k--parameter
in the {\it concordance model} \cite{lahav}, we can infer the corresponding value for 
a fitted $\Lambda-$parameter in the same diagram, since for negligible $k-$parameter,
$\Omega^{\CD}_{\cal Q} + \Omega^{\CD}_{\cal R} = 1 - \Omega^\CD_m \sim
\Omega^F_{\Lambda}$, where with the upper index $F$ we refer to a `fitted' 
$\Lambda-$parameter.

Another feature in the following diagram are arrows that illustrate the time--evolution
of the respective parameters. Again,
the models corresponding to $\Omega_{m}^{\CD}>1$ are not analyzed here
(among them are `Big--Crunch--models', {\it cf.}~Eq~(\ref{phaseevol}); 
their dynamics depends strongly on initial conditions).
There are attractors and repellors\footnote{Compare the analyses 
in \cite{ehlersrindler} and \cite{sota:RG}.
Due to the proposed correspondence to a scalar field cosmology we can directly
use the results on scaling properties of the scalar field investigated by 
\cite{copeland:scaling} and \cite{amendola:scaling}. 
In this context, the determination of equations of state from
similarity symmetries also provides an interesting tool \cite{generalscaling}.
A full--scale investigation of a dynamical systems analysis is not provided here.} 
in this diagram.
Most notably, for the cases of interest to us that feature a late--time acceleration, 
`Friedmannian' states are repellors, i.e. `near--Friedmannian' states
evolve away down to the attractor solution $\Omega^\CD_m = 0$, where backreaction-- (or
curvature--) domination is completed.

Let us specifically discuss the various cases. We denote values in the solution plane by
($q^\CD ; \Omega^\CD_m$) and concentrate in the following only on {\it expanding} 
universe models. We write:
\begin{equation}
\label{phaseevol}
q_{\CD}(a_{\CD})\;=\;\frac{1}{2}\frac{1+c_{1}a_{\CD}^{n+3}}{1+c_{2}a_{\CD}^{n+3}}
\;\;\;;\;\;\;
\Omega_{m}^{\CD}\;=\;\frac{1}{1+c_{2}a_{\CD}^{n+3}}\;\;,
\end{equation}
with $c_{1}=4rc_{2}/(1+r)$, $c_{2}=(1-\Omega_{m}^{\initial
  \CD})/\Omega_{m}^{\initial \CD}$, and $a_{\CD}$ being an increasing function of time.
These formulae are helpful to determine the evolution of a particular
solution in the phase plane ($q^\CD ; \Omega^\CD_m$).
We learn from Figure~\ref{fig:cosmicstates} that no model corresponding to {\it Case C}
can produce acceleration.
According to Eqs.~(\ref{phaseevol}), scaling solutions in the sectors corresponding to negative potentials (case E) and 
pit--type potentials (case D) are attracted towards the Einstein--de Sitter model
($1/2 ; 1$). In all the other
cases (A, B and C), the Einstein--de Sitter model appears as a repellor, and
the attractors are located on the line  ($q^\CD = 2r/(1+r) ; \Omega^{\CD}_m=0$)
for $r\in ]-1;0]$. 
Each point of the straight line $r=1/3$, that corresponds to a renormalized
Einstein--de Sitter scenario, is a fixed point.

Hence, the Einstein--de Sitter model is a
 saddle point for the scaling dynamics and small inhomogeneities with 
${\cal Q}_{\CD}>0$ should make the system evolve away from it.
The sign of ${\cal Q}_{\CD}$ is important: for all the
models corresponding to $r>0$ or $r<-1$, that is the cases C,D and E, which cannot
produce accelerated expansion, we have ${\cal Q}_{\CD}<0$. In other words, the
kinematical backreaction is dominated by shear fluctuations, {\it cf.} 
Eq.~(\ref{eq:Q-GR}). This does not necessarily mean that the
universe model is regionally (on the scale $\CD$) anisotropic, because in these 
cases kinematical fluctuations decay strongly.  On the other hand,
cases A and B that could be responsible for an accelerated expansion
correspond to ${\cal Q}_{\CD}>0$ and have subdominant shear fluctuations. 
Therefore, these models can be regionally almost 
isotropic, although kinematical fluctuations have strong influence.

Finally, recall that we have ${\cal R}_{\initial\CD}<0$ for $r>-1$ and ${\cal  R}_{\initial\CD}>0$ 
for $r<-1$.

\subsection{Construction of a realistic model: estimation of parameters and initial conditions}

The scalar field behavior (i.e. the late--time behavior of the cosmological model in the cases
$r<1/3$) essentially depends on the value of the backreaction parameter $r$ that 
describes the ratio between kinematical backreaction and averaged curvature 
(or, in the scalar field language, the 
ratio between the kinetic energy of the field and its potential energy).
We are going to estimate this ratio from observational constraints on the
equation of state.
Before we do so, let us summarize the conditions relevant for a late--time behavior featuring
`volume acceleration', i.e. ${\ddot a}_\CD > 0$.

\subsubsection{Acceleration conditions}
\label{subsubsect:accelerationconditions}

The condition for an accelerating patch $\CD$ (which we are going to take as large as our 
observable Universe) follows from the averaged Raychaudhuri equation
(\ref{averageraychaudhuri}) \cite{kolbetal}, \cite{buchert:static}:
\begin{equation}
\label{accelerationcondition1}
{\cal Q}_\CD \;>\;4\pi G\langle\varrho\rangle_\CD \;\;,
\end{equation}
implying with (\ref{omega}) and (\ref{deceleration}):
\begin{equation}
\label{accelerationcondition2}
-\Omega^\CD_{\cal Q}\;>\;\frac{\Omega_m^\CD }{4} \;\;\;;\;\;\;q^\CD \;<\;0\;\;,
\end{equation}
and, for the class of scaling solutions (\ref{neqpsolution}) ($\Omega_m^\CD > 0$, 
$\Omega^\CD_{\Lambda} =0$, and the definition (\ref{gammaR})):
\begin{equation}
\label{accelerationcondition3}
\fl
\Omega^\CD_{\cal Q} = r \,\Omega^\CD_{\cal R} \;\,,\,{\rm i.e.,}\;\;-r\,\gamma^\CD_{\CR m}
\;>\;\frac{1}{4} \;\;\;\Leftrightarrow\;\;\;q^\CD = \Omega_m^\CD \left(\frac{1}{2} + 2 r \,
\gamma^\CD_{\CR m} \right) \;<\;0\;\;,
\end{equation}
This condition is met, if (now inserting the solution (\ref{gammaR}) and the relation to the
Friedmannian curvature parameter (\ref{gammaK})):
\begin{equation}
\label{accelerationcondition4}
-r\,\gamma^\CD_{\CR m} = -\frac{r}{(1+r)} \gamma_{k m}^{\initial\CD} \,
a_\CD^{(n+3)}\;>\;\frac{1}{4}  \;\;.
\end{equation}
A realistic model would meet this condition at some time in the evolution leading thereafter to
the observed acceleration value, 
i.e. ideally at the epoch around the time of structure formation solving the 
coincidence problem.

From what has been said above, such realistic cases require $r < -1/3$, i.e., $n >0$;
in the limiting case $r=-1/3$ $(n=0)$ (which is the solution mimicking a cosmological constant) the 
condition (\ref{accelerationcondition4}) reads 
$\gamma^{\initial\CD}_{k m}\,a_\CD^3 \;>\;1/2$, and with $\Omega^\CD_k = 
\Omega^{\initial\CD}_k
\,a_\CD^{-2}$ and $\Omega^\CD_m = \Omega^{\initial\CD}_m\,a_\CD^{-3}$ we find:
\begin{equation}
\label{accelerationcondition5}
r=-\frac{1}{3}\;\;,\;\;n=0\;\;:\;\;\gamma^\CD_{k m} \,a_\CD^2 \;>\;\frac{1}{2}\;\,,\;\;{\rm i.e.}
\;,\;\;
\gamma^{\now\CD}_{k m} \;>\;  \frac{1}{2 a_{\now\CD}^2} \;\;.
\end{equation}  
The marginal case $n=-3$ in the exponent of the volume scale factor 
(\ref{accelerationcondition4})
plays a particular role that we shall explain more in detail in
Appendix A. In the context of the acceleration conditions we can gain a better understanding of
the cases $n>-3$ by the following remark. In this marginal case the averaged density and the
kinematical backreaction have identical decay rates with respect to the volume scale factor
in an expanding universe model. This means that, in order to get a positive acceleration at the
present time, already the initial data must satisfy the conditions (\ref{accelerationcondition1})
and (\ref{accelerationcondition2}). The necessary initial value for kinematical backreaction 
is then large and suggests that we are looking at a region $\cal D$ that is close to satisfy 
the condition needed for the stationarity of the cosmos. 
Note, however, that the strict single--scaling solution 
$n=-3$ ($r=1/3$) does not admit 
acceleration, {\it cf.} Eq.~(\ref{decelerationscaling}), but a superimposed 
regional fluctuation would admit acceleration (or deceleration) ({\it cf.} Appendix A). 

In the cases $n>-3$, the averaged density decays faster than kinematical backreaction;
hence, to attain sufficient acceleration today, the model needs the less magnitude of kinematical
backreaction the weaker its decay given in terms of $n$. Let us take a case 
close to $n=-2$ (the case $n=-2$ degenerates to $r=0$, {\it cf.} Eq.~(\ref{neqpsolution})),
then kinematical backreaction can be three orders of magnitude weaker initially, if the
scale factor advanced to a value of $a_{\now\CD} =1000$ today.
This remark makes clear that the solution sector $n>-2$ ($r<0$) contains solutions that 
can potentially explain the Dark Energy problem even when starting with small expansion
fluctuations at the CMB (Cosmic Microwave Background)
epoch. Moreover, it contains solutions which also solve the
coincidence problem, although a more natural solution would not be an exact scaling solution,
but one that would inject more backreaction at the formation epoch of structure. 
However, for models with $\Omega^\CD_m < 1$, 
{\it cf.} Eq.~(\ref{decelerationscaling}), we have to go to values of $n \sim 0$ 
($r \sim -1/3$) in order to find {\it sufficient} acceleration.

It should be emphasized that the interesting sector $n>-2$ is {\it not} what we could find 
in a weakly perturbed FRW (Friedmann Robertson Walker) model.
These states rely on a strong coupling of kinematical fluctuations to the
averaged scalar curvature of the universe model. Kinematical backreaction can only decay
at such weak rates (or even grow for $r<-1/3$), 
if the time--evolution of the averaged scalar curvature largely deviates from the time--evolution
of a constant curvature model; intuitively speaking, averaged fluctuations are strongly 
supplied by the `curvature energy reservoir'. 

\subsubsection{Observational constraints on parameters}
\label{subsect:observationalconstraints}

Recall that the envisaged class of single--scaling solutions implies that our parameter
choices are unambiguous: we only have to specify an initial condition, say
$\Omega^{\initial\CD}_{\cal R}$ or $\Omega^{\initial\CD}_{m}$, 
the backreaction parameter $r$, and the value of the 
volume scale factor today $a_{\now \CD}$.

The following estimates are done on the assumption that the effective model is based on
solutions for a matter density source and a morphon field that is realized by the particular
class of scaling solutions discussed in this paper. 
We focus on the effect of the morphon field for vanishing
cosmological constant, and would like to demonstrate that an observer who is
using a Friedmannian `template universe model'  would interprete this effect by 
a cosmological constant {\it today}. 
Thus, we are {\it forcing} the effective evolution of
the volume scale factor to match with that of a Friedmannian model with $\Lambda$
at initial and final time.
However, 
all of our parameters are evolving according to the `best--fit' scaling solution
in the averaged {\it inhomogeneous} model. 
In particular, this implies that the time--derivatives
of the volume scale factor evolve very differently compared with a standard Friedmannian
model.  Thanks to the
existing constraints on the standard Friedmannian models, as for example Cold Dark Matter
models with a Dark Energy component that has a constant equation of
state, this procedure reduces the number of parameters that we have to
estimate. Indeed, we then only have to determine $r$ and one value for the initial
data, e.g. $\Omega^{\initial\CD}_{\cal R}$.

We emphasize that the interpretation of observational data and the resulting constraints 
strongly depend on model assumptions, i.e. it is commonly assumed that a standard 
$\Lambda CDM$ model is the correct one.
This implies, in particular, that we are not constraining the parameters of the 
averaged inhomogeneous model by observations reinterpreted within the inhomogeneous 
cosmology. For example,
since the spatial curvature is not constant, the formulae for
angular diameter and luminosity distances cannot be taken as the FRW ones.
Furthermore, we point out that this reinterpretation is challenging, for all the other 
observational predictions that are based 
on a perturbative approach, like large--scale structure characteristics, must be reconsidered.
It is not obvious, and this work together with others (e.g. \cite{rasanen:model} and references
therein) provides plausible counter arguments, 
that the {\it late Universe} could be described by a perturbed FRW model,
even if smoothed over large scales. 

In this sense, our analysis is a demonstration of what the
observed values of the standard cosmological parameters 
would imply for the averaged quantities.

With these assumptions the observer with a `Friedmannian template' then faces the 
following relation {\it today}:
$$
\Omega_{\cal R}^{\now \CD} + \Omega_{\cal  Q}^{\now \CD}\;=\;
\Omega^{F}_{\Lambda}+\Omega^{\now\CD}_k\;\;,
$$
where the latter corresponds to the `biased' interpretation of the
true dynamics. The parameter $r$ is fully specified by
the energy content of the Universe today, since by Hamilton's constraint
$\Omega_{\cal R}^{\now \CD} + \Omega_{\cal  Q}^{\now \CD}=
1-\Omega_m^{\now\CD}$.
 
Directly following from the relation (\ref{constraininglambda}) we have:
\begin{equation}
\label{r_vsomegas}
r=-\frac{n+2}{n+6} \quad{\rm with}\quad 
(n+2)\,=\,\frac{\ln \left(1+\gamma^{\now\CD}_{\Lambda k}
\right)}{\ln a_{\now \CD}}\;\;;\;\;
\gamma^{\now\CD}_{\Lambda k}:= 
\frac{\Omega^{\now\CD}_{\Lambda}}{\Omega^{\now \CD}_k}\;\;.
\end{equation}
Note that, again by Hamilton's constraint, the condition 
$1+ \gamma^{\now\CD}_{\Lambda k} >0$
implies $(1-\Omega^{\now\CD}_m)/\Omega^{\now\CD}_k >0$, which allows for two 
cases: a positive  curvature today ($\Omega^{\now \CD}_{k}$ negative) with 
$\Omega^{\now\CD}_m >1$, or a negative curvature today with $\Omega^{\now\CD}_m <1$.
For our purpose of fitting the inhomogeneous model to a Friedmannian `template' with
cosmological constant, we choose the latter option.
Furthermore, for $\Omega^F_{\Lambda}>0$, there is still the possibility that 
$\Omega^{F}_{\Lambda}<-\Omega^{\now \CD}_{k}$. 
This last condition is clearly not satisfied in the late--time Universe, 
so in the following, we will restrict the analysis to $\Omega^{\now \CD}_{k}>0$.

\begin{figure}
\begin{center}
\includegraphics[width=10cm]{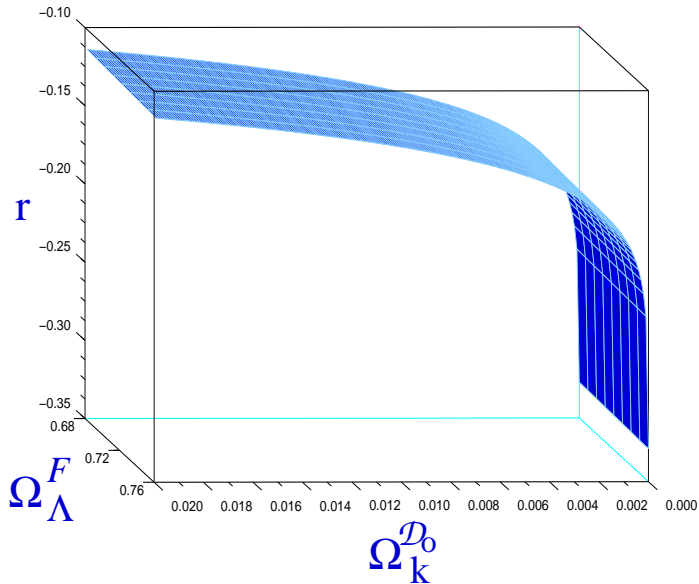}
\caption{\label{fig:r_vs_param}
The backreaction parameter $r$ is shown as a function of 
$\Omega^{\now\CD}_{k}$ and $\Omega^{F}_{\Lambda}$, as the 
effect of backreaction would
be interpreted in a Friedmannian `template', with $a_{\now \CD}\sim 1000$.}
\end{center}
\end{figure}

Giving initial data from WMAP \cite{WMAP3},
 $\Omega^{F}_{\Lambda} \sim 0.72$ and $\Omega^{\now\CD}_{m} \sim 0.28$,
and taking $a_{\now\CD} \sim 1000$,
Eq.~(\ref{r_vsomegas}) determines the value of $n$ to be close to but slightly larger than
$0$, i.e. $r$ slightly below $r =-1/3$ pointing to a phantom quintessence.
However, taking into account a variation of the initial data, we 
detect a large sensitivity of the precise value for $r$ to the initial data in our 
scaling solutions. This is shown in 
Figure~\ref{fig:r_vs_param}. We therefore employ an orthogonal observational constraint
on the Dark Energy component below.

Figure \ref{fig:r_vs_param} shows $r$ as a function of $\Omega^{\now\CD}_{k}$ 
and $\Omega^{F}_{\Lambda}$ for the bounds given by WMAP
third year data \cite{WMAP3}: $\Omega^{\now \CD}_{k}\in ]0;0.006]$ and
  $\Omega^{F}_{\Lambda}=0.72\pm0.04$, and for $a_{\now \CD}=1000$ that
 roughly  corresponds to a range of solutions integrated from approximately the epoch of 
the CMB (Cosmic Microwave Background).

We infer that when $\Omega^{\now \CD}_{k}$ tends to $0$, $r$ tends to
$-1$, and to $0$ when $\Omega^{\now \CD}_{k}$ tends to $+\infty$. 
Moreover, we notice that, while
increasing $\Omega^{\now \CD}_{k}$ from zero, the morphon field is first a
phantom field, then a standard quintessence field.
Finally, $r$ is almost insensitive to $\Omega^{F}_{\Lambda}$, and it
depends strongly on  $\Omega^{\now \CD}_{k}$ only in the case when 
$\Omega^{\now \CD}_{k}$ is very small. (Note that the latter is a `pathology' of our
single--scaling ansatz; this behavior could be cured by superimposing a constant--curvature
solution.) 

\subsubsection{Constraining $r$ by supernovae observations}
\label{subsect:concrete}

The preceding considerations giving $r$ in terms of $\Omega_{\Lambda}^{F}$ and
$\Omega_{k}^{\now \CD}$ (or in terms of $\Omega_{\Lambda}^{F}$ and
$\Omega_{m}^{\now \CD}$ by replacing $\Omega_{k}^{\now \CD}$ with
$1-\Omega_{m}^{\now\CD}\Omega_{\Lambda}^{F}$ in Eq.~(\ref{r_vsomegas}))
can be used to determine $r$.
We now estimate $r$ through the Dark Energy equation of state that best matches
supernovae observations. Then, we
use relation (\ref{r_vsomegas}) to compute the resulting value of
$\Omega_{k}^{\now\CD}$, in order to check the consistency with our
fitting model. 

There is a simple way to estimate the backreaction parameter $r$, by requiring that
the equation of state for Dark Energy $w^\CD_{\Phi}$, {\it cf.} 
Eq.~(\ref{eosmorphon1}),  matches the one
inferred from supernovae observations, denoted by $w_{SN1a}$, provided this
one is constant. Then:
$$
r=\frac{1}{3}\frac{1+3w_{SN1a}}{1-w_{SN1a}}\;\;.
$$
Taking into account the last SNLS best fit for a flat Friedmann model sourced by
Dark Energy with a constant equation of state
$w_{SN1a}\sim-1.02$, $\Omega^{\now\CD}_{\Lambda}=0.73$ 
and $\Omega^{\now\CD}_{m}=0.27$
\cite{SNLS}, we find $r\sim-0.34$, again suggesting a phantom quintessence.
This value does not depend sensitively  on variations in $w_{SN1a}$, and is therefore
a more robust estimate compared with our previous one.

Once this ratio is fixed, we can find the values of the initial data. 
We assume that, today,
$\Omega_{\cal R}^{\now\CD}+\Omega_{\cal Q}^{\now\CD}=
\Omega^{F}_{\Lambda}$ and find, e.g. for the initial ratio 
of the curvature parameter to the density parameter: 
$$
\gamma^{\initial\CD}_{\CR m}=
\frac{-{\cal R}_{\initial\CD}}{16\pi G\langle\varrho\rangle_{\initial\CD}}
=\frac{\Omega^{F}_{\Lambda}}{(1+r)\Omega^{\now\CD}_{m}\,
a_{\now\CD}^{(n+3)}}\;\;,
$$
which, at the CMB epoch, approximately setting $a_{\now\CD}$ 
to the value $a_{\now\CD}\sim 1000$, is 
$\gamma^{\initial\CD}_{\CR m} \sim 2.7\, 10^{-9}$.

Here, we can check the consistency of the SNLS fitting with the curvature assumption. 
Indeed, in the SNLS fitting, we assumed that $\Omega^{\now \CD}_k=0$, and, 
using the inferred value for $r$, $r\sim-0.34$ in Equation (\ref{r_vsomegas}), 
we find that $\Omega^{\now \CD}_k \sim 5\,10^{-7}\,\Omega^{F}_\Lambda$, 
in accordance with the assumption $\Omega^{\now \CD}_k=0$. This shows that
$r$ as determined only through the Dark Energy equation of state is compatible
without further assumptions.

\begin{figure}
\begin{center}
\includegraphics[width=16.5cm,height=16.5cm,angle=270]{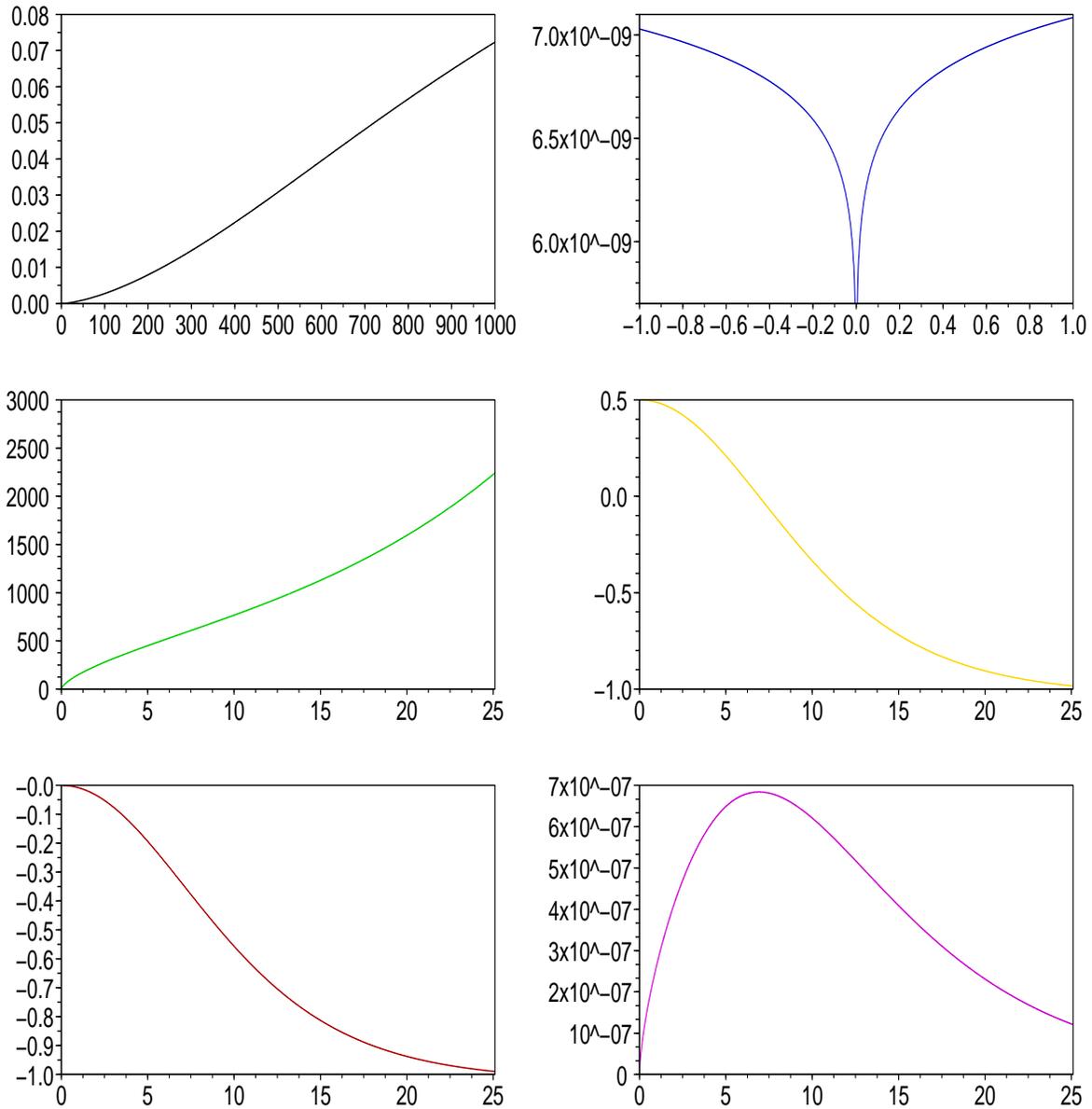}
\caption{\label{fig:SNLS}
Upper left: scalar field $\Phi_\CD$ as a function of the volume scale factor
$a_\CD$, in units of $2\sqrt{\pi G}$. Upper right:
Potential $U(\Phi_\CD)$, in GeV m$^{-3}$. Central
left: volume scale factor as a function of time in Gyr. Central right:
deceleration parameter as a function of time in Gyr. Lower left: 
the `cosmic equation of state' $w_{\rm eff}$ 
as a function of time in Gyr. Lower right: `Friedmannian
curvature parameter' as a function of time in Gyr, featuring a maximum at the 
onset of acceleration.
All solutions are calculated for the SNLS best fit model.}
\end{center}
\end{figure}

\begin{figure}
\begin{center}
\includegraphics[width=16cm]{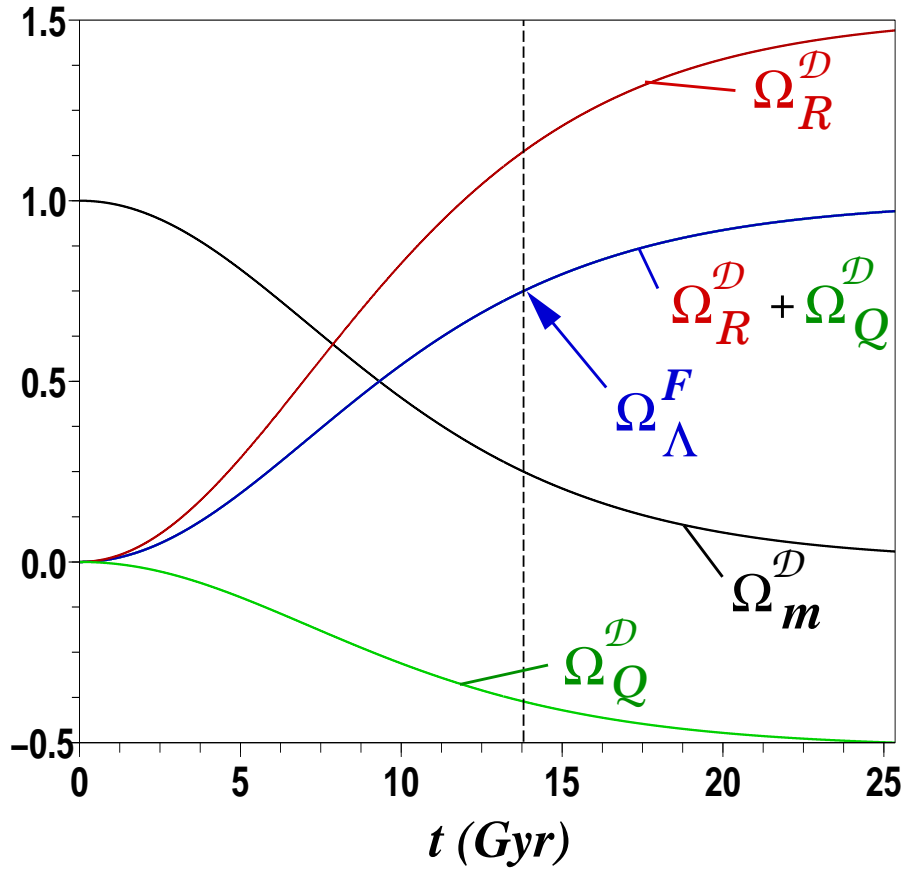}
\caption{\label{fig:omegasSNLS}
The dimensionless cosmological parameters are shown as a function of time. 
While $\Omega^{\CD}_{m}$ decays from $1$ to the present value $\sim 0.27$, the sum
$\Omega^{\CD}_{\cal Q}+ \Omega^\CD_{\cal R}$ grows from $0$ to the value
$\sim 0.73$, interpreted as the cosmological constant parameter today,
$\Omega^F_{\Lambda}$, in a Friedmannian `template model'. Kinematical 
backreaction ${\cal Q}_\CD$ is positive (dominant expansion fluctuations)
and grows slightly, implying a negative 
backreaction parameter $\Omega^{\CD}_{\cal Q}$ that becomes more negative. 
This growth is supplied by the negative averaged curvature that evolves strongly, i.e. the
positive curvature parameter $\Omega^{\CD}_{\cal R}$ grows from almost $0$ to
a value larger than $1$ at the present epoch. This behavior of the `physical' 
curvature parameter should be compared with the evolution of the `Friedmannian 
curvature parameter' displayed in the last figure.
All solutions are calculated for the SNLS best fit model. In this model
the parameters ($\Omega^{\CD}_{m}; \Omega^{\CD}_{\cal Q};\Omega^\CD_{\cal R}$)
evolve towards their attractor ($0;-0,515;+1,515$), where backreaction-- and
curvature--domination is completed. Note that the starting values differ from 
($\Omega^{\initial\CD}_{m}=1; \Omega^{\initial\CD}_{\cal Q}=0;
\Omega^{\initial\CD}_{\cal R}=0$); if they would not, then the model evolves without backreaction
as the standard model, in which the latter initial data set is maintained in the course of evolution.
The standard model is a repellor for the scenario shown.
}
\end{center}
\end{figure}

If we define an `effective redshift' through the volume scale factor as in 
Friedmann cosmology,
we can derive the effective redshift at which the expansion
accelerates. It corresponds to a `cosmic equation of state' for matter plus
backreaction $w^{\cal D}_{\rm eff} \sim -1/3$. Inserting 
(\ref{eosmorphon1plusmatter}) into this relation, we find an acceleration
scale factor and an effective acceleration redshift:
$$
a_{\CD}^{\rm acc}=
\left(\, -1 /(4r\gamma_{\CR m}^{\initial\CD})\,\right)^{\frac{1+r}{1-3r}}\;\;
\sim 569\;\;\;\Rightarrow\;\;\;
z_{\rm eff}^{\rm acc}=\frac{a_{\now\CD}}{a_{\CD}}-1\;\sim \;0.76\;\;.
$$
The scalar field behavior and the potential corresponding to this model, as well as
the time--evolution of various cosmological parameters,  are
presented in Figs.~\ref{fig:SNLS} and \ref{fig:omegasSNLS}.

\clearpage

\section{Concluding Remarks and Outlook}

Our proposal of a mean field description of backreaction effects through a minimally
coupled `morphon field', does not only provide a rephrasing of the kinematics of backreaction
in terms of a scalar field cosmology, but it also justifies existence of the latter due to the fact that
we identify averaged inhomogeneities in Einstein gravity as the underlying fundamental physics.
Thus far, such a well--defined link was missing; alternatively, it is thought and there exist some plausible hints
that the low--energy scale in some contemporary  particle physics models 
would provide this link (e.g. \cite{nilles}, \cite{polchinski}, \cite{karthauser}), 
i.e. that the scalar field emerges from the dynamical degrees of freedom 
stemming from extra dimensions (`moduli fields').

The set of spatially averaged equations together with the mean field description of
kinematical backreaction by a morphon field shows, in particular, that the framework of
Friedmann--type equations is very robust. We understand now that an effective and
scale--dependent Friedmannian framework is applicable with the surprising new input
that, besides averaged matter sources, also a scalar field component emerges and has 
not to be invented.

\subsection{Summary}

We have exploited the proposed correspondence by investigating a family of scaling solutions
of the backreaction terms. The study of scaling solutions is well--advanced in research work on 
scalar field models,
and our study allows to translate those results into the backreaction context.
Here, our discussion was not exhaustive. We concentrated on exploring the solution space of 
inhomogeneous cosmologies with the help of scaling solutions with a `minimal' parameterization.
Therefore, as a next step, this parameterization could be expanded and congruences with other work
on scalar field models could be worked out. Note that an obvious such expansion would analyze
superimposed scaling solutions that could be modeled either again by a single effective scalar
field, or else by multiple scalar fields \cite{copeland:multiple}, \cite{collinucci}.
One example of a superposition of scaling solutions is provided in Appendix A.
Furthermore, observational constraints set on quintessence or phantom quintessence
models have direct relevance to constraints that have to be set on morphon fields 
\cite{maartens:phantom}. However, observations have to be reinterpreted within the
inhomogeneous cosmology underlying the morphon dynamics.

In making the correspondence concrete 
we have not touched the question whether realistic dynamical models of the inhomogeneous
Universe would comply with the class of averaged models that we singled out as `realistic case
studies'.
For the scaling solutions the scalar field correspondence revealed, that the constancy of the 
fraction of kinetic to potential energy of the morphon field relates to averaged models that 
are driven by strong coupling between averaged 3--Ricci curvature and kinematical fluctuations,
the constancy implying their direct proportionality. While in the scalar field context this 
assumption is often employed, here we get an interesting picture of what this represents. 
The key for a physical interpretation of the proposed correspondence is the
equivalence of the integrability condition (\ref{integrability})
and the Klein--Gordon equation (\ref{kleingordon}):
$$
a_\CD^{-6}\partial_{t}\left(a_\CD^{6}{\cal Q}_\CD\right)+
a_\CD^{-2}\partial_{t}\left(a_\CD^{2}\average{\CR}\right)\,=\,0\;\;,
\;\;\Leftrightarrow\;\;
{\ddot\Phi}_\CD + 3 H_{\cal D}{\dot\Phi}_\CD + 
\epsilon\frac{\partial}{\partial \Phi_\CD}U(\Phi_\CD)\;=\;0\;\;.
$$
Although the resulting picture may appear complex, it can be structured by looking at 
a single player: the averaged 3--Ricci curvature.
If it is positive, it represents  an energy reservoir, stored in a negative potential in the 
correspondence. If this reservoir exceeds the `virial energy', then the excess energy is
converted (in the scaling solutions directly)
into an excess of kinetic energy, i.e. kinematical backreaction. The averaged curvature then
decays faster than the `Friedmannian' curvature. 
On the other hand, if it is negative, then the same logic applies: the averaged curvature
consequently becomes stronger negative compared with the `Friedmannian' curvature.
Solutions of the 
Dark Energy problem, where a large positive value of backreaction is needed, will have to 
make strong use of this conversion, while `near--Friedmannian' models don't.
A Newtonian or post--Newtonian approach suppresses these degrees of freedom by
freezing the averaged curvature to the `Friedmannian' value (compare an attempt to work
with the Friedmannian curvature parameter that, of course, needs an 
extra scalar field as a source for Dark Energy \cite{franca}). 
From these remarks we understand the importance of a strongly evolving averaged 3--Ricci
curvature, if the attempt to explain Dark Energy through backreaction should be successful.
In this line there is clear support for the importance of a strongly evolving 3--Ricci 
curvature from the Lema\^\i tre--Tolman--Bondi solution, see \cite{rasanen:LTB},
and in particular the recent work \cite{curvatureLTB} and references to other
papers on this solution therein. Recently, R\"as\"anen
\cite{rasanen:model} provided an illustrative example and a comprehensive discussion
of the physics of backreaction--driven accelerated expansion. 

Weakly perturbed Friedmann models have a special status if placed into the full
solution space of the averaged inhomogeneous models: only if  the averaged fluctuations
{\it decouple} from the averaged curvature, i.e. if they evolve independently, then the
averaged curvature evolves as a constant `Friedmannian' curvature, and 
fluctuations decay in proportion to the square of the inverse volume. 
Hints that point to the likely existence of a curvature--fluctuation--coupling
come again from averaging the LTB solution \cite{nambu} (see also \cite{chuang}) and
other global solutions \cite{buchert:static} 
that all show an extra term in the averaged scalar curvature.
In these examples the extra term  evolves in proportion to the inverse volume,  
hence deviates from a 
`Friedmannian' curvature and implies maintainance of large kinematical fluctuations that 
decay  only in proportion to the inverse volume.  While this particular 
behavior of the backreaction terms still requires
large backreaction to start with \cite{buchert:darkenergy}, 
e.g. a feature of globally stationary solutions \cite{buchert:static}, a stronger evolution of averaged
curvature, i.e. injecting more curvature energy into kinematical fluctuations, would allow to
start with `near--Friedmannian' initial conditions and still explain current observations.
An example for the latter possibility has been analyzed in this paper.
This possibility can be interpreted such that there is some hope to find
enough backreaction by starting with almost FRW initial data, as suggested by 
\cite{kolbetal}, 
although here the conversion of curvature energy into backreaction must be very 
efficient; a more moderate evolution into backreaction--dominated phases 
would result, if we already start with `out--of--equilibrium' initial data.

In this line we have also pointed out, and this is worth stressing again,  
that the `Friedmannian' curvature parameter $k_{\initial\CD}$ 
is in general unrelated to the physical
averaged scalar curvature, it is an integration constant and actually an integral of motion
\cite{buchert:static}: the dynamics of the averaged physical curvature is not represented
by and, in most examples, deviates substantially from it. The restriction to a 
`Friedmannian' evolution of the averaged curvature singles out the special case where the curvature--fluctuation--coupling
is {\it strictly absent}.

Summarizing these thoughts we can say that there are two conceivable scenarios that
remove the need for Dark Energy: first, a `soft scenario' that was discussed in detail
in \cite{buchert:static}. Here we already have an initial global state with strong expansion
fluctuations, so that regionally a moderate evolution of backreaction and averaged 
curvature suffices to explain the observed acceleration. 
It was, however, pointed out that such models imply
paradigmatic changes, and observational data have to be reinterpreted in order to put 
firm constraints on ${\cal Q}_\CD$ at the CMB epoch. Second, there is a `hard scenario'
(implied by the suggestion of Kolb et al. \cite{kolbetal})
that literally creates enough backreaction out of `nothing', 
{\it cf.} Eq.~(\ref{earlydarkenergy}). In our example 
of a particular scaling solution a phantom
quintessence scenario arises that complies with the strong energy condition and also
with  constraints in accord with those already put on the
standard model. Consequently, this scenario needs strong evolution of backreaction and 
averaged curvature as seen best for the dimensionless cosmological parameters 
in Fig.~\ref{fig:omegasSNLS}.  The present work has shown that the `hard' version 
can be made consistent with the framework of the averaged Einstein equations.

\subsection{Outlook}

There are obviously a number of possible routes for generalizing the scalar field correspondence.
Let us briefly discuss some of them. 

Spatial averages are scale--dependent, since we integrate over
a given compact volume of the space sections. In this paper we left the domain--dependence 
untouched, all of our considerations were focussing on a fixed scale. However, the 
scalar field correspondence holds on every scale, which is not only reminiscent of, but also 
physically a manifestation of renormalizable quantities. In this respect the morphon
analogues of quintessence are different from standard quintessence models.
In this context we could
`Ricci flow' the averages to averages on a constant curvature geometry \cite{klingon}, or we
could study other renormalization group methods to control the scale--dependence, 
e.g. \cite{calzetta}, \cite{gaite:RG}, \cite{carfora:focker}. 

A possibly fruitful investigation would consider string--motivated gravitational theories
and, employing the proposed correspondence, would aim at determining the scalar field theory
from the higher--dimensional geometry of extrinsic and intrinsic curvature. For this end
one would have to derive the averaged equations for the extended spacetime. 
Brane world cosmologies could be analyzed in the spirit of this correspondence.
Not only in this context it will be interesting to understand, 
in which cases a non--minimally coupled morphon field would arise. 

Furthermore, there are other interesting strategies of a more technical nature that,
however, widen the physical context of applications. 
For example,  the introduction of a complex scalar field, which is necessary to access
solutions with $\Omega^\CD_m >1$ with a positive averaged 3--Ricci curvature; 
the discussion of globally stationary cosmologies in
\cite{buchert:static} shows that a stationary cosmos obeys these conditions at early times,
and later the averaged scalar curvature decreases fast until it becomes negative, while
maintaining large kinematical backreaction.  Another example is
the averaging on a different foliation of spacetime, i.e. 
introducing coordinate degrees of freedom. The latter is actually necessary if matter sources other
than `irrotational dust' are considered; it results in an extra coupling of the scalar field to the
matter sources stemming from a backreaction term due to the pressure gradient that is
absent in the present paper \cite{buchert:grgfluid}. 
This is the subject of a follow--up work that will also allow access to
realistic models of backreaction--driven inflation \cite{khlopov}. 
Here, the present investigation in principle allows to model
inflationary scenarios too by translating typical characteristics of inflaton fields into 
corresponding morphon fields, however, within the restricted cases of `dust matter' or
`vacuum'.

Since $\Phi_\CD$ models the inhomogeneous `vacuum part' of the
sources,  there might be an interesting connection to the energy budget of 
gravitational waves that
could be fruitfully exploited \cite{dautcourt}, and, 
since $\Phi_\CD$ represents a `mean field' of fluctuations,
also a connection to statistical thermodynamics is implicit. 
In order to establish this latter link, however,
we have to note that ${\cal Q}_\CD$ models the averaged spatial variance of
extrinsic curvature and so far
not `fluctuations' in the thermodynamical sense.
Another intimate, but on the level of a `dust matter model' formal, 
relation to effective thermodynamic models is suggested in the case
of imperfect fluid models. They imply non--equilibrium effects that in turn can be associated
with a `friction coefficient' proportional to the `cosmic equation of state'  
$p_{\rm eff}/\varrho_{\rm eff}$. Thermodynamic arguments that were used 
in \cite{zimdahl,schwarz} within the imperfect fluid picture lead to further
possibilities of interpreting the different scaling regimes, e.g. by employing the second law of 
thermodynamics the authors of \cite{zimdahl,schwarz} would conclude $p_{\rm eff} < 0$, 
and thus ${\cal Q}_\CD > 1/3 \langle \CR\rangle_{\CD}$, i.e.
$r > 1/3$ for positive averaged scalar curvature and $r < 1/3$ for negative one. 
Since the first case does not give rise to acceleration one would be led to 
conclude that the scalar curvature has to be negative on average, a conclusion shared by
other work (in Appendix A we also reach this conclusion
and provide references). 

A final comment related to scalar field perturbations, often investigated in a
post--Newtonian setting, is in order. In those perturbative approaches, 
the scalar degree of freedom that emerges is a combination of matter and
metric inhomogeneities. Within the well--developed standard 
perturbation theory \cite{mukhanov}, it is natural to first restrict the analysis to  
a tight range around a FRW background in order to calculate backreaction effects
\cite{futamase1,futamase2}, \cite{abramo1,abramo2}.
The amplitude of scalar perturbations, but also their  derivatives
and, hence,  all curvature terms have to be small \cite{wald}. 
Besides considerations of long--wavelength perturbations (e.g., \cite{branden1}),
the analysis of sub--horizon perturbations  with the aim to explain Dark Energy 
in a perturbative setting is the focus of many recent research papers 
and, here, we do not enter into the details of these works. 
We wish to point out two aspects that emerged from the present investigation, and which
may be relevant for future strategies in a perturbation analysis.

First, we address the averaging issue: a post--Newtonian approximation may
(and in most cases of cosmological relevance will) 
be adequate {\it locally} or piecewise on a small range of scales. However,  as soon as we are 
looking at integral properties, i.e. integrating out a wider range of scales, its applicability 
should be verified.  In standard perturbation theory spatial averages are taken 
with respect to a `background observer'. Since a major player in the mechanism that 
can produce large kinematical fluctuations is the averaged 3--Ricci curvature, we have to
be careful in relating Euclidean `background averages' to the Riemannian volume averages 
that govern the dynamics of the averaged cosmology.

Second, in the present approach we do not
specify a metric of the space sections; the formulation and the correspondence holds for
arbitrary 3--metrics. Also, when we spoke about a `scale--dependent Friedmannian model', 
we referred to the kinematics of the volume scale factor, we did not refer to the FRW metric.
We can, however, specify a spatial metric
to  establish a dynamical model.  Here, we think that the metric setting must allow for 
large deviations of the 3--Ricci curvature from the constant Friedmannian curvature. 
As a `rule of thumb' (another was recently given in \cite{parry} for super--Hubble 
perturbations), we may say: 
{\it if the averaged scalar curvature evolves at or near the
constant curvature model, then there is no hope to model cases that lead to enough
late--time acceleration.} 

A future strategy related to perturbation theory could be motivated by 
Newtonian cosmological models for structure formation:
in the Newtonian framework, an Eulerian perturbation theory does not provide
access to the highly non--linear regime. Instead, the Lagrangian point of view offers
a way to move with the largely perturbed fluid. 
A relativistic Lagrangian  perturbative approach is currently worked out aiming at 
generalizing the Newtonian work \cite{bks} that has employed the exact averaged equations to 
construct a non--perturbative model for inhomogeneities out of perturbatively
calculated fluctuations.
A Lagrangian perturbation scheme itself does not include
non--perturbative features, that may be needed in this context; non--perturbative effects
have been recently discussed in the Newtonian framework \cite{buchert:nonperturbative}.

\vspace{6pt}

All these efforts aim at constructing a generic evolution model.  
The morphon field is an effective description without any perturbative assumption. But, one would
like to establish the underlying inhomogeneous dynamics in order to understand, to what the realistic case studies
that `explain away' the Dark Energy problem actually correspond.

\newpage

\subsection*{\it Acknowledgements:}

{\footnotesize
TB and JL  acknowledge support by the Sonderforschungsbereich SFB 375 
`Astroparticle physics' by the German science foundation DFG, JL during a visit
to ASC Munich. TB would like to thank 
the Observatory of Meudon, Paris, and the 
University of Bielefeld, for hospitality and support. In particular, he would like to thank
Dominik Schwarz for his invitation to temporarily hold a chair in Bielefeld university,
where an excellent working environment and stimulating discussions with collegues
in the physics department made this stay most enjoyable.
TB and JL would like to thank Matthew Parry and Herbert Wagner for 
fruitful discussions during the preparation stage of this work;
TB would like to also thank Misao Sasaki for encouraging conversations during a 
Dark Energy workshop held in Munich, and Mauro Carfora, Martin Kerscher, 
Slava Mukhanov, Syksy R\"as\"anen, Varun Sahni and Dominik Schwarz for valuable 
remarks and discussions.}

\renewcommand{\theequation}{A.\arabic{equation}}
\setcounter{equation}{1}  
\section*{Appendix A: a relation to Friedmannian fitting models and the globally stationary
solution}

A particular scaling behavior of the solutions (\ref{neqpsolution}), in which kinematical
backreaction and averaged scalar curvature are proportional, can be exploited to define
a mapping of the solutions (\ref{neqpsolution}) for the particular case $r=1/3$, i.e. $n=-3$, to a 
Friedmannian cosmology.

Before we explain this mapping let us recall that the case $r=1/3$ also arises for the
stationarity condition, required in \cite{buchert:static} for the global scale (extending the
domain $\CD$ to the whole (compact) manifold $\Sigma$, and setting $\Lambda = 0$):
\begin{equation}
\label{globalstationarity}
3\frac{{\ddot a}_{\Sigma}}{a_{\Sigma}} = 
-\frac{4\pi G\langle\varrho\rangle_{\initial\Sigma} +{\cal Q}_{\initial\Sigma}}{a_{\Sigma}^3}
  = 0\;\;.
\end{equation}
This condition implies either a globally static model or a globally stationary model featuring
the solution:
\begin{equation}
\label{stationaryS}
{\cal Q}_{\Sigma} \;=\; \frac{{\cal Q}_{\initial\Sigma}}{a_{\Sigma}^{3}}\;\;\;;\;\;\;
\gaverage{\CR} \;=\;  \frac{{\cal R}_{\initial\Sigma}-3{\cal Q}_{\initial\Sigma}}
{a_{\Sigma}^{2}} \;+\;\frac{3{\cal Q}_{\initial\Sigma}}{a_{\Sigma}^{3}}  \;\;,
\end{equation}
where $\;H_{\Sigma} = {\cal C}/a_{\Sigma}\;$ with:
$\;-6{\cal C}^2 = {\cal R}_{\initial\Sigma}-3{\cal Q}_{\initial\Sigma}$.

\vspace{3pt}

The relation to a Friedmannian `fitting model' arises by noting that, in order to obtain the 
above solution, we do not need to assume the stationarity condition  (\ref{globalstationarity}).
Indeed, inserting the scaling solutions  (\ref{neqpsolution}) into the averaged 
equations (\ref{averageraychaudhuri}) and (\ref{averagehamilton}) on any given domain
$\CD$, and superimposing a constant curvature solution to the averaged curvature, 
we obtain (restricting again attention to $\Lambda =0$):
\begin{equation}
\fl
3\frac{{\ddot a}_\CD}{a_\CD} + 
\frac{4\pi G\langle\varrho\rangle_{\initial\CD} -{\cal Q}_{\initial\CD}}{a_{\CD}^3}  = 0\;\;\;
;\;\;\;\label{averageF}
\left( \frac{{\dot a}_\CD}{a_\CD}\right)^2 - 
\frac{16\pi G\langle\varrho\rangle_{\initial\CD}- (\frac{1}{r}+1) 
{\cal Q}_{\initial\CD}}{6 a_{\CD}^3}-\frac{{\cal C}^2}{a_\CD^2}=0\;\;.
\end{equation}
We find that, for the special case $r=1/3$, we can simply redefine the initial constants,
\begin{equation}
\label{redefine}
\fl
k^F_\CD : =-{\cal C}^2\;\;;\;\; 
4\pi G \langle\varrho\rangle_{\initial\CD}^F :=4\pi G\langle\varrho\rangle_{\initial\CD}-
{\cal Q}_{\initial\CD}\;\;;\;\;\Omega^\CD_{mF} + \Omega^\CD_k:=\Omega^\CD_m + \Omega^\CD_{\cal Q} + 
\Omega^\CD_{\cal R} = 1\,,
\end{equation}
where $\Omega^\CD_{mF}$ and $\Omega^\CD_k$ denote the resulting dimensionless functionals 
which we are going to use as Friedmannian `fitting parameters', so that Eqs.~(\ref{averageF}) assume the
form of a constant--(negative) curvature Friedmannian model.

Let us exemplify this correspondence. 
Setting  $k^F_\CD \sim 0$ in accord with the current observational results, we have 
$\Omega^\CD_{mF} = 1$ throughout the evolution, 
i.e. a scale--dependent Einstein--de Sitter model:
\begin{equation}
\label{einsteindesitterD}
3\frac{{\ddot a}_\CD}{a_\CD} + 
\frac{4\pi G\langle\varrho\rangle_{\initial\CD}^F}{3 a_{\CD}^3}  = 0\;\;\;
;\;\;\;
\left( \frac{{\dot a}_\CD}{a_\CD}\right)^2 - 
\frac{8\pi G\langle\varrho\rangle_{\initial\CD}^F}{ a_{\CD}^3}=0\;\;.
\end{equation}
Suppose now that we `fit' a standard Einstein--de Sitter model on some given scale $\CD$
to observational data, we would be in the position to evaluate the physical `parameters'
on that scale to be $\Omega^\CD_m = 1 - 4\Omega^\CD_{\cal Q}$, i.e. for 
$\Omega^\CD_m \sim 1/3$ today, we would conclude that there must be backreaction at work
and it should be positive (negative kinematical backreaction mimicking a `kinematical dark matter' 
source),
$\Omega^\CD_{\cal Q} \sim 1/6$, and that the (physical) 
curvature parameter would be positive too (negative averaged
curvature), $\Omega^\CD_{\cal R} = 3 \Omega^\CD_{\cal Q} \sim 1/2$.
We emphasise that the Friedmannian curvature parameter is assumed to vanish, which 
demonstrates that it has nothing to do with the (evolving) averaged scalar curvature of
the inhomogeneous model. 

The above procedure exemplifies the possibility of constructing a (non--naive) Friedmannian
fitting model. It, however, assumes that, regionally, the model obeys an Einstein--de Sitter
kinematics (unrelated to an underlying FRW metric), 
and $r= 1/3$ is `typical' for the regional Universe. 
Both are not in accord with what we expect. We would  `fit' a Friedmannian model 
{\it with} a cosmological constant, which is the
currently held view of the `concordance model',
\begin{equation}
\label{fitlambda}
\Omega^\CD_{mF} + \Omega^\CD_{k}+\Omega^\CD_{\Lambda} \;=\;1\;\;, 
\end{equation}
then we would have to superimpose a solution $\Omega^\CD_{{\cal Q}2}$
with, e.g., $r = -1/3$ to the above solution (this could still be interpreted as a deviation 
from a representative volume of a global model with $r=1/3$).
This implies, 
with $\Omega^\CD_{mF} = \Omega^\CD_m +4\Omega^\CD_{{\cal Q}1} \sim 1/3\;;\;
\Omega^\CD_{\Lambda} \sim 2/3$, 
$\Omega^\CD_{k} \sim 0$, and
\begin{equation}
\label{Q2}
\Omega^\CD_m + \Omega^\CD_{{\cal Q}1} + \Omega^\CD_{{\cal Q}2} +
\Omega^\CD_{{\CR}1} + \Omega^\CD_{{\CR}2} \;=\;1\;\;,
\end{equation}
that 
\begin{equation}
\label{Lambda2}
\Omega^\CD_{\Lambda} = -2 \Omega^\CD_{{\cal Q}2} \;\sim \frac{2}{3}\;\;. 
\end{equation}
We would find a positive backreaction with $\Omega^\CD_{{\cal Q}2} \sim -1/3$
(modeling now Dark Energy),
and a negative averaged scalar curvature with $\Omega^\CD_{{\CR}2} \sim 1$,
indicating that the regional Universe should correspond to a `void' within a global model
with $\Omega^{\Sigma}_{{\CR}1} \sim 1/2$.
That our regional Universe could correspond to a regional `void' has been discussed in a number
of other papers, e.g.  
 \cite{tomita1,tomita2,tomita3}, \cite{linde2}, \cite{wiltshire}, \cite{alnes:LTB},
\cite{moffat}.

This example also demonstrates that we can construct cosmologies with different
properties on global and regional scales by superimposing scaling solutions.

\end{document}